\documentclass[12pt,a4paper]{article}
\usepackage[utf8]{inputenc}
\usepackage{epsf}
\usepackage{latexsym,amssymb,euscript}
\usepackage[dvips]{graphicx}
\usepackage{calc}
\usepackage[numbers,sort&compress]{natbib}
\usepackage{amsmath}
\usepackage{slashed}
\usepackage{booktabs}
\usepackage{hyperref}
\usepackage{braket}
\usepackage{chngcntr}
\usepackage{bbold}
\usepackage{graphics}
\usepackage{graphicx}
\usepackage{caption}
\usepackage{subcaption}
\usepackage{pdfpages}
\usepackage[titletoc]{appendix}
\graphicspath{{./figures/}}
\hypersetup{
 linktocpage = true,
 urlcolor = purple,
 colorlinks = true,
 linkcolor = purple,
 anchorcolor = purple,
 citecolor = purple,
 pdfstartview = {XYZ null null 1.25} 
           }
\usepackage[left=2cm, right=2cm]{geometry}
\usepackage{pstricks}
\usepackage{color}
\usepackage{float}
\usepackage{academicons}
\definecolor{orcidlogocol}{HTML}{A6CE39}
\usepackage{fancyhdr}
\DeclareRobustCommand\textastdbl{%
  \leavevmode
  {\sbox0{\ddag}%
   \ooalign{\raisebox{\ht0-\height}{*}\cr
            \raisebox{\depth-\dp0}{\scalebox{1}[-1]{*}}\cr}%
  }%
}

\newcommand{\tcite}[1]{~\cite{#1}}

\newcommand{\orcidADB}{\href{https://orcid.org/0000-0002-6114-7044}{\includegraphics[scale=0.1]{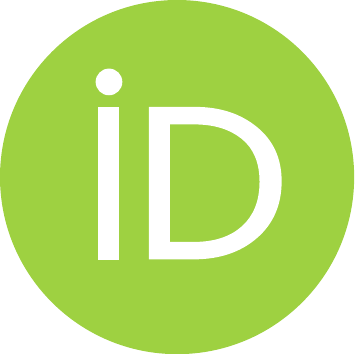}}}

\newcommand{\orcidFGC}{\href{https://orcid.org/0000-0003-3299-2203}{\includegraphics[scale=0.1]{figures/logo-orcid.pdf}}}

\newcommand{\orcidDYI}{\href{https://orcid.org/0000-0001-5701-4364}{\includegraphics[scale=0.1]{figures/logo-orcid.pdf}}}

\newcommand{\orcidAP}{\href{https://orcid.org/0000-0001-8984-3036}{\includegraphics[scale=0.1]{figures/logo-orcid.pdf}}}

\newcommand{\orcidWS}{\href{https://orcid.org/0000-0002-9764-3138}{\includegraphics[scale=0.1]{figures/logo-orcid.pdf}}}

\newcommand{\orcidAS}{\href{https://orcid.org/0000-0001-5247-8442}{\includegraphics[scale=0.1]{figures/logo-orcid.pdf}}}

\begin{document}

\begin{titlepage}

\newgeometry{bottom=0pt}

\begin{center}
  {\LARGE \bf Exclusive production of $\rho$-mesons \\ \vspace{0.0cm} in high-energy factorization \\ \vspace{0.225cm} at HERA and EIC}
\end{center}

\vskip 0.3cm

\centerline{
A.D.~Bolognino$^{1,2,*}$ \orcidADB,
F.G.~Celiberto$^{3,4,5,\dagger}$ \orcidFGC,
D.Yu.~Ivanov$^{6,\ddagger}$ \orcidDYI,
} \vspace{.25cm}
\centerline{
A.~Papa$^{1,2,\S}$ \orcidAP,
W.~Sch\"afer$^{7\P}$ \orcidWS,
and A.~Szczurek$^{7,8\textastdbl}$ \orcidAS
}

\vskip .4cm

\centerline{${}^1$ {\sl Dipartimento di Fisica, Universit\`a della Calabria,}}
\centerline{\sl I-87036 Arcavacata di Rende, Cosenza, Italy}
\vskip .2cm
\centerline{${}^2$ {\sl Istituto Nazionale di Fisica Nucleare, Gruppo collegato di Cosenza,}}
\centerline{\sl I-87036 Arcavacata di Rende, Cosenza, Italy}
\vskip .2cm
\centerline{${}^3$ {\sl European Centre for Theoretical Studies in Nuclear Physics and Related Areas (ECT*),}}
\centerline{\sl I-38123 Villazzano, Trento, Italy}
\vskip .2cm
\centerline{${}^4$ {\sl Fondazione Bruno Kessler (FBK), %}}
%\centerline{\sl
I-38123 Povo, Trento, Italy} }
\vskip .2cm
\centerline{${}^5$ {\sl INFN-TIFPA Trento Institute of Fundamental Physics and Applications,}}
\centerline{\sl I-38123 Povo, Trento, Italy}
\vskip .2cm
\centerline{${}^6$ {\sl Sobolev Institute of Mathematics, 630090 Novosibirsk, Russia}}
\vskip .2cm
\centerline{${}^7$ {\sl Institute of Nuclear Physics Polish Academy of Sciences,}}
\centerline{\sl ul. Radzikowskiego 152, PL-31-342, Krak\'ow, Poland}
\vskip .2cm
\centerline{${}^8$ {\sl College of Natural Sciences, Institute of Physics, University of Rzesz\'ow,}}
\centerline{\sl ul. Pigonia 1, PL-35-310 Rzesz\'ow, Poland}
\vskip .2cm

\vskip .4cm

\begin{abstract}
\vspace{0.50cm}
\hrule \vspace{0.75cm}
We study cross sections for the exclusive diffractive leptoproduction of $\rho$-mesons, $\gamma^*~p~\to~\rho~p$, within the framework of high-energy factorization. Cross sections for longitudinally and transversally polarized mesons are shown.
We employ a wide variety of unintegrated gluon distributions available in the literature and compare to HERA data. The resulting cross sections strongly depend on the choice of unintegrated gluon distribution.
We also present predictions for the proton target in the kinematics of the Brookhaven EIC.
\vspace{0.75cm} \hrule
\vspace{0.75cm}
%{
% \setlength{\parindent}{0pt}
% \textsc{Keywords}: QCD phenomenology, exclusive processes, high-energy factorization, hadronic structure
%}
\end{abstract}

\vskip -.6cm

$^{*}${\it e-mail}:
\href{mailto:ad.bolognino@unical.it}{ad.bolognino@unical.it}

$^{\dagger}${\it e-mail}:
\href{mailto:fceliberto@ectstar.eu}{fceliberto@ectstar.eu}

$^{\ddagger}${\it e-mail}:
\href{mailto:d-ivanov@math.nsc.ru}{d-ivanov@math.nsc.ru}

$^{\S}${\it e-mail}:
\href{mailto:alessandro.papa@fis.unical.it}{alessandro.papa@fis.unical.it}

$^{\P}${\it e-mail}:
\href{mailto:wolfgang.schafer@ifj.edu.pl}{wolfgang.schafer@ifj.edu.pl}

$^{\textastdbl}${\it e-mail}:
\href{mailto:antoni.szczurek@ifj.edu.pl}{antoni.szczurek@ifj.edu.pl}

\end{titlepage}

\restoregeometry

%----------------------------
\section{Introduction}
%----------------------------
\label{introd}

The diffractive exclusive electroproduction of vector mesons has been a very active field of study at the HERA collider at DESY. Especially it has served as a testbed for the perturbative QCD description of the Pomeron exchange mechanism which drives  
diffractive and elastic processes~\cite{Ivanov:2004ax, Donnachie:2002en}.
New precise data, albeit in a different kinematic regime, are expected  to be taken at the Brookhaven Electron-Ion Collider (EIC) in a not too distant future~\cite{Accardi:2012qut,AbdulKhalek:2021gbh}.
In this work we focus on the electroproduction of $\rho$-mesons, {\it i.e.} on the process $\gamma^* p \to \rho p$ at large virtuality $Q^2$ of the photon and large $\gamma^* p$-system center-of-mass energy $W$, with $ {x \simeq Q^2/W^2 \ll 1}$.

\begin{figure}[t]
\centering
\includegraphics[width=0.50\textwidth]{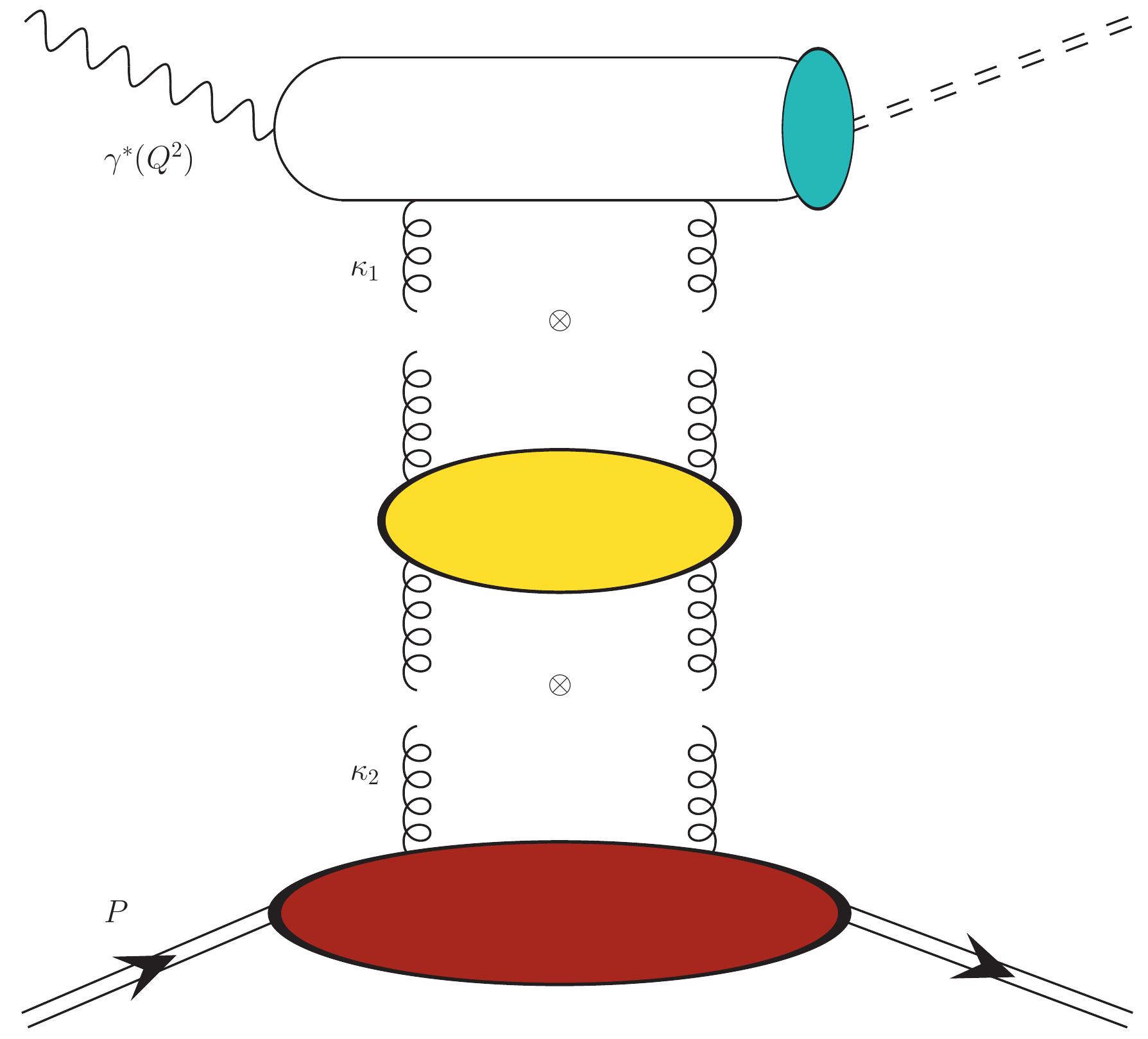}
\caption{Diagrammatic representation of the amplitude for the exclusive emission of a $\rho$ meson in high-energy factorization. The off-shell impact factor (upper part) is built up as a collinear convolution between the hard factor for the photon splitting to a dipole and the non-perturbative $\rho$-meson DA (sea-green blob). The UGD is given by the convolution of the BFKL gluon Green's function (yellow blob) and the non-perturbative proton impact factor (red blob).}
\label{fig:process_rho}
\end{figure}

The large photon virtuality $Q^2$ justifies the use of perturbation theory.
A diagrammatic representation of the perturbative QCD factorization structure of the forward amplitude is shown in Fig.~\ref{fig:process_rho}.
At large $Q^2$ the transverse internal motion of quark and antiquark in the $\rho$-meson can be integrated out, and all information is contained in the distribution amplitude (DA).
The transverse momenta of gluons in the proton must be fully taken into account by utilizing the unintegrated gluon distribution (UGD). In the BFKL-approach the UGD is obtained by convoluting the BFKL two-gluon Green's function with the proton impact factor (IF).

The IF for the transition $\gamma^*(\lambda_\gamma) \to \rho (\lambda_\rho)$ depends on polarizations of photon and vector mesons. For the longitudinal polarizations $\lambda_\gamma = \lambda_\rho = 0$ the dominance of small dipoles is evident, and the standard leading-twist distribution amplitude appears. In the case of the transverse polarizations the perturbative QCD factorization remains valid, but higher-twist DAs are needed~\cite{Anikin:2009bf,Anikin:2011sa}.
The difference between the impact factors makes the polarization dependence of diffractive production a sensitive probe of the UGD~\cite{Bolognino:2018mlw,Bolognino:2018rhb}.
Previously, for the case of $\rho$-meson production ratios of the forward amplitudes have been calculated in Ref.~\cite{Bolognino:2018rhb} and compared to data from HERA. For the case of $\phi$ mesons the effect of a finite strange quark mass on cross sections has been discussed in Ref.~\cite{Bolognino:2019pba}, where also relations of the so-called genuine higher-twist DAs to weighted integrals over light-front wave functions have been given.

In this work, we wish to extend the efforts of~\cite{Bolognino:2018mlw,Bolognino:2018rhb,Bolognino:2019pba} to the calculation of polarized $\rho$-meson production cross sections (as opposed to ratios of amplitudes) for a variety of unintegrated gluon distributions, and compare the results to HERA data. 
We will also give predictions for the kinematics relevant for the Electron-Ion Collider EIC.

%-----------------------------------------------------------------
\section{Polarized $\rho$-meson leptoproduction}
%-----------------------------------------------------------------
\label{process}
The H1 and ZEUS collaborations have provided extended analyses of the helicity structure in the hard exclusive production of the $\rho$ meson in $ep$ collisions through the subprocess 
\begin{equation}
\label{process_rho}
\gamma^*(\lambda_\gamma)p\rightarrow \rho (\lambda_\rho)p\,.
\end{equation}
Here $\lambda_\rho$ and $\lambda_\gamma$ represent the meson and photon
helicities, respectively, and can take the values 0 (longitudinal polarization)
and $\pm 1$ (transverse polarizations). The helicity amplitudes 
$T_{\lambda_\rho \lambda_\gamma}$ extracted at HERA~\cite{Aaron:2009xp} exhibit the
hierarchy
\begin{equation}
\label{hierarchy}
T_{00} \gg T_{11} \gg T_{10} \gg T_{01} \gg T_{-11} \, ,
\end{equation}
that follows from the dominance of a  small-size 
dipole scattering mechanism, as discussed first in Ref.~\cite{Ivanov:1998gk}, see also~\cite{Kuraev:1998ht}. As we previously mentioned, the H1 and ZEUS collaborations have analyzed data in different ranges of $Q^2$
and $W$. In what follows we will refer only to the H1 ranges~\eqref{H1range},
\begin{equation}
\label{H1range}
\begin{split}
2.5\,\text{\rm GeV$^2$} < Q^2 <60 \,\text{\rm GeV$^2$}\,,\\
35\, \text{GeV} < W < 180\,\text{GeV}\,,
\end{split}
\end{equation}
illustrating the framework to probe the $\rho$-meson leptoproduction as a way to test the UGDs~\cite{Bolognino:2018mlw,Bolognino:2019bko,Celiberto:2019slj} where polarized cross sections will be the key observables.

%---------------------------------------
\subsection{Theoretical setup}
%---------------------------------------
\label{theory}
In the high-energy regime, $s\equiv W^2\gg Q^2\gg\Lambda_{\text{QCD}}^2$, which
implies small $x = Q^2/W^2$, the forward helicity amplitude for the $\rho$-meson electroproduction can be written, in high-energy factorization (also known as $k_T$-factorization), as
the convolution of the $\gamma^*\rightarrow \rho$ IF,
$\Phi^{\gamma^*(\lambda_\gamma)\rightarrow\rho(\lambda_\rho)}(\kappa^2,Q^2)$,
with the UGD, ${\cal F}(x,\kappa^2)$. Its expression reads
\begin{equation}
\label{amplitude}
T_{\lambda_\rho\lambda_\gamma}(s,Q^2) = \frac{is}{(2\pi)^2}\int \dfrac{d^2\kappa}
{(\kappa^2)^2}\Phi^{\gamma^*(\lambda_\gamma)\rightarrow\rho(\lambda_\rho)}(\kappa^2,Q^2)
{\cal F}(x,\kappa^2),\quad \text{$x=\frac{Q^2}{s}$}\,.
\end{equation}

Defining $\alpha = \frac{\kappa^2}{Q^2}$ and $B =~2\pi \alpha_s
\frac{e}{\sqrt{2}}f_\rho$, the expression for the IFs takes the
following forms (see Ref.\tcite{Anikin:2009bf} for the derivation):

\begin{itemize}
	
	\item longitudinal case
	\begin{equation}
	\label{Phi_LL}
	\Phi^{\gamma_L\rightarrow\rho_L}(\kappa,Q;\mu^2) = 2 B\frac{\sqrt{N_c^2-1}}{Q\,N_c}
	\int^{1}_{0}dy\, \varphi_1(y;\mu^2)\left(\frac{\alpha}{\alpha + y\bar{y}}\right)
	\,,
	\end{equation}
	where $N_c$ denotes the number of colors and
	$\varphi_1(y;\mu^2)$ is the twist-2 DA which, up to
	the second order in the expansion in Gegenbauer polynomials,
	reads\tcite{Ball:1998sk}
	\begin{equation}
	\label{phi}
	\varphi_1(y; \mu^2) = 6y\bar{y}\left(1+a_2(\mu^2)\frac{3}{2}
	\left(5 (y-\bar{y})^2-1\right)\right)\,;
	\end{equation}
	
	\item transverse case
	\[
	\Phi^{\gamma_T\rightarrow\rho_T}(\alpha,Q;\mu^2)=
	\dfrac{(\epsilon_{\gamma}\cdot\epsilon^{*}_{\rho}) \, 2 B m_{\rho}
		\sqrt{N_c^2-1}}{2 N_c Q^2}
	\]
	\begin{equation}
	\times\left\{-\int^{1}_{0} dy \frac{\alpha (\alpha +2 y \bar{y})}{y\bar{y}
		(\alpha+y\bar{y})^2}\right. \left[(y-\bar{y})\varphi_1^T(y;\mu^2)
	+ \varphi_A^T(y;\mu^2)\right]
	\label{Phi_TT}
	\end{equation}
	\[
	+\int^{1}_{0}dy_2\int^{y_2}_{0}dy_1 \frac{y_1\bar{y}_1\alpha}
	{\alpha+y_1\bar{y}_1}\left[\frac{2-N_c/C_F}{\alpha(y_1+\bar{y}_2)
		+y_1\bar{y}_2}
	-\frac{N_c}{C_F}\frac{1}{y_2 \alpha+y_1(y_2-y_1)}\right]
	\]
	\[
	\left.\times M(y_1,y_2;\mu^2)-\!\int^{1}_{0}dy_2\!\int^{y_2}_{0}dy_1\! \left[\frac{2+N_c/C_F}{\bar{y}_1}
	+\frac{y_1}{\alpha+y_1\bar{y}_1}\right.\right.
	\]
	\[
	\!\left.\left.\times\left(\frac{(2-N_c/C_F)
		y_1\alpha}{\alpha(y_1+\bar{y}_2)+y_1\bar{y}_2}-2\right)-\frac{N_c}{C_F}\frac{(y_2-y_1)\bar{y}_2}{\bar{y}_1}\right.\right.
	\]
	\[
	\left.\left.\times\frac{1}{\alpha\bar{y}_1+(y_2-y_1)\bar{y}_2}\right]\!S(y_1,y_2;\mu^2)\right\}\,,
	\]
	where $C_F=\frac{N_c^2-1}{2N_c}$, while the functions $M(y_1,y_2;\mu^2)$
	and $S(y_1,y_2;\mu^2)$ are defined in Eqs.~(12)-(13) of
	Ref.\tcite{Anikin:2011sa}
	and are combinations of the twist-3 DAs $B(y_1,y_2;\mu^2)$ and
	$D(y_1,y_2;\mu^2)$ (see Ref.~\cite{Ball:1998sk}), given by
	\begin{align}
	\label{BD}
	B(y_1,y_2;\mu^2) & =-5040 y_1 \bar{y}_2 (y_1-\bar{y}_2)
        (y_2-y_1) \notag \; ,\\
	D(y_1,y_2;\mu^2) & =-360 y_1\bar{y}_2(y_2-y_1)
	\left(1+\frac{\omega^{A}_{\{1,0\}}(\mu^2)}{2}\left(7\left(y_2-y_1\right)-3\right)
	\right)\,.
	\end{align}
	
\end{itemize}

In Eqs.~\eqref{phi} and~\eqref{BD} the functional dependence of
$a_2$, $\omega^{A}_{\{1,0\}}$, $\zeta^{A}_{3\rho}$, and $\zeta^{V}_{3\rho}$ on the
factorization scale $\mu^2$ can be determined from the corresponding known
evolution equations\tcite{Ball:1998sk}, using some suitable initial condition
at a scale $\mu_0$. 

Note that the $\kappa$-dependence of the IFs is different in the cases of
longitudinal and transverse polarizations and this poses a strong constraint
on the $\kappa$-dependence of the UGD in the HERA energy range. The main
point will be to demonstrate, considering different models of UGD, that
the uncertainties of the theoretical description do not prevent us from
some, at least qualitative, conclusions about the shape of 
the UGD in $\kappa$.

The DAs $\varphi^T_1(y;\mu^2)$ and $\varphi^T_A(y;\mu^2)$ in Eq.~\eqref{Phi_TT}
encompass both genuine twist-3 and Wandzura-Wilczek~(WW)
contributions\tcite{Anikin:2011sa,Ball:1998sk}. The former are related to
$B(y_1,y_2;\mu^2)$ and $D(y_1,y_2;\mu^2)$; the latter are those obtained
in the approximation in which $B(y_1,y_2;\mu^2)=D(y_1,y_2;\mu^2)=0$, and in
this case read\footnote{For asymptotic form of the twist-2 DA,
	$\varphi_1(y)=\varphi_1^{\text{as}}(y)=6y\bar y$, these equations give
	$\varphi^{T\;WW,\;\text{as}}_A(y)=-3/2y\bar y$ and $\varphi^{T\;WW,\;\text{as}}_1(y)=
	-3/2 y\bar y (2y -1)$.}
\[
\label{WWT}
\varphi^{T\;WW}_A(y;\mu^2)= \frac{1}{2}\left[-\bar y
\int_0^{y}\,dv \frac{\varphi_1(v;\mu^2)}{\bar v} -
y \int_{y}^1\,dv \frac{\varphi_1(v;\mu^2)}{v}   \right]\;,
\]
\begin{equation}
\varphi^{T\;WW}_1(y;\mu^2)= \frac{1}{2}\left[
-\bar y \int_0^{y}\,dv \frac{\varphi_1(v;\mu^2)}{\bar v} +
y \int_{y}^1\,dv \frac{\varphi_1(v;\mu^2)}{v}   \right]\;.
\end{equation}

\begin{figure}[t]
\centering

\includegraphics[scale=0.55,clip]{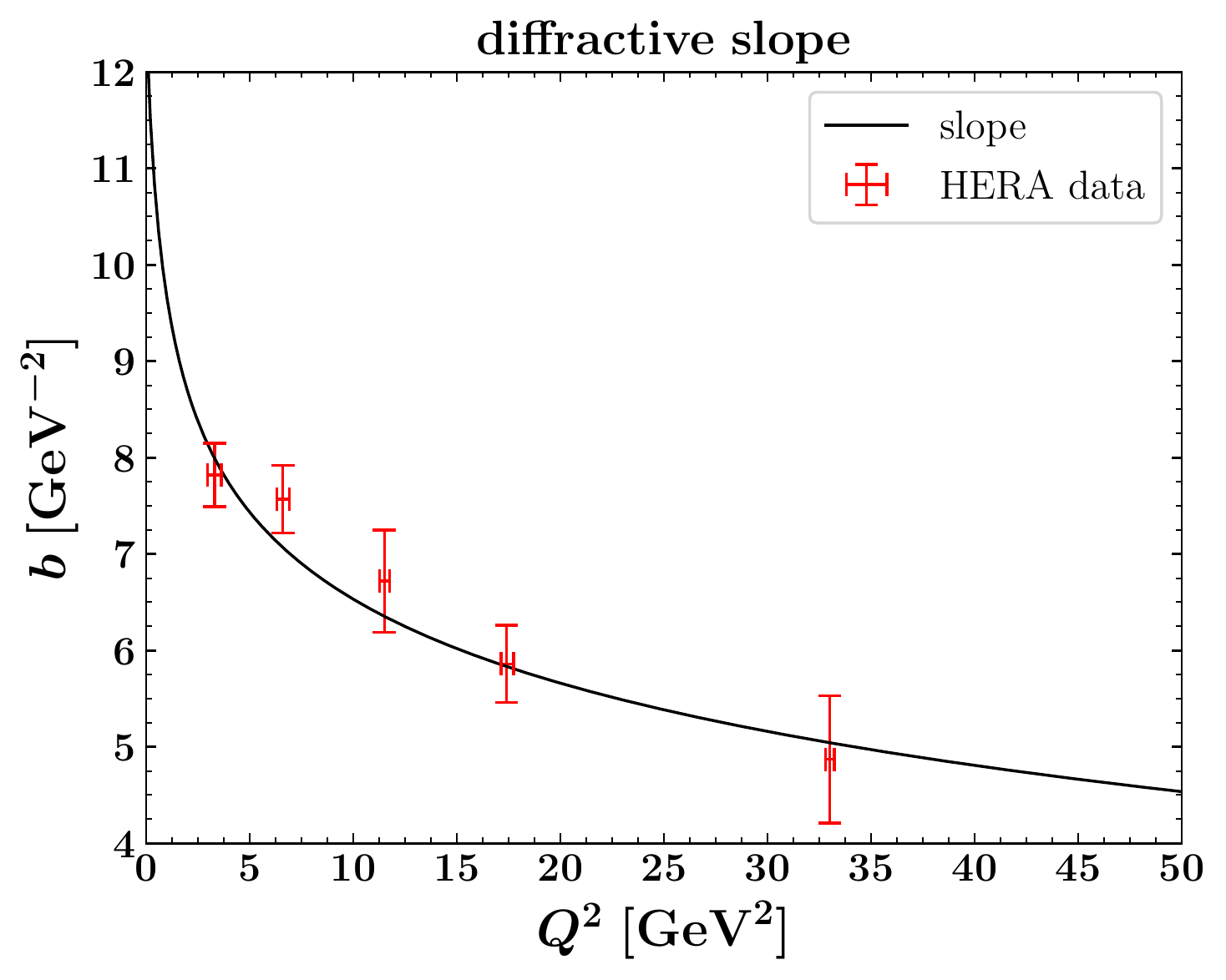}
\caption{$Q^2$-dependence of the diffractive slope for the exclusive $\rho$-meson leptoproduction~(Eq.~\eqref{slope_B_rho}) compared to the H1 data from Ref.~\cite{Aaron:2009xp}.}
\label{fig:slope}
\end{figure}

The other interesting point will be the extension of information,
collected by the helicity-amplitude analysis, reachable from the
calculation of the cross section. As a matter of fact, the expressions
for the polarized cross sections $\sigma_L$ and $\sigma_T$, calculated
using 
Eqs.~\eqref{Phi_LL}, \eqref{phi} in Eq.~\eqref{amplitude} and 
Eqs.~\eqref{Phi_TT}, \eqref{BD} in Eq.~\eqref{amplitude}, respectively, are
\begin{equation}
\label{sigL_rho}
\sigma_L\,(\gamma^*\,p \rightarrow \rho\,p) = \frac{1}{16 \pi b(Q^2)}\left|\frac{T_{00}(s, t = 0)}{W^2}\right|^2\,,
\end{equation}
\begin{equation}
\label{sigT_rho}
\sigma_T\,(\gamma^*\,p \rightarrow \rho\,p) = \frac{1}{16 \pi b(Q^2)}\left|\frac{T_{11}(s, t = 0)}{W^2}\right|^2\,,
\end{equation}
where $b(Q^2)$ in Eqs.~\eqref{sigL_rho} and~\eqref{sigT_rho} is the diffraction slope, for which we adopt the parametrization\tcite{Nemchik:1997xb}:
\begin{equation}
\label{slope_B_rho}
b(Q^2) = \beta_0 - \beta_1\,\log\left[\frac{Q^2+m_\rho^2}{m^2_{J/\psi}}\right]+\frac{\beta_2}{Q^2+m_\rho^2}\,,
\end{equation}
fixing the values of the constants as follows: $\beta_0 = 6.5$ GeV$^{-2}$, $\beta_1 = 1.2$  GeV$^{-2}$ and $\beta_2 = 1.6$.
In Fig.~\ref{fig:slope}, we compare the parametrization of Eq.~\eqref{slope_B_rho} with the data of the H1-collaboration from Ref.~\cite{Aaron:2009xp}.
Above, we took into account only the helicity conserving amplitudes. These dominate the diffractive peak at small $t$, 
and hence also the integrated cross section of interest.
While the impact factors for the helicity flip transitions $\gamma^*(T)\to \rho(L)$ with $\Delta \lambda = \pm 1$ and $\gamma^*(T)\to \rho(T')$ with  $\Delta \lambda = \pm 2$,
are available within the higher-twist factorization approach \cite{Anikin:2009bf,Anikin:2011sa},
a discussion of the $t-$dependent cross section or the polarization density matrix of $\rho$-mesons would require also a generalization to off-forward UGDs and goes beyond the scope of this work. 

The advantage of considering polarized cross sections relies on the
possibility to constrain not only the shape and the behavior of
results, but also the \emph{normalization}, now essential and
which is clearly irrelevant for the evaluation of the helicity-amplitude
ratio $T_{11}/T_{00}$. Although all UGDs described in Section~\ref{UGD}
present fixed values for their parameters, the ABIPSW model, one of the
first UGD parametrization adopted in phenomenological
analysis~\cite{Forshaw}, was defined up to the overall normalization. As regards the helicity-amplitude ratio $T_{11}/T_{00}$,
results for the ABIPSW UGD model show a fair agreement with
data\tcite{Anikin:2011sa}, this suggesting that the shape, controlled by
the $M$ parameter (see Eq.~\eqref{ABIPSW}), has been already
guessed. The best value of the normalization parameter
can be obtained via a simple global fit to experimental data of both polarized cross sections, $\sigma_L$ and $\sigma_T$.

%---------------------------------------------------------------
\section{UGD models used in the present study}
%---------------------------------------------------------------
\label{UGD}

We have considered a selection of several models of UGD, without
pretension to exhaustive coverage, but with the aim of comparing
(sometimes radically) different approaches. We refer the reader to
the original papers for details on the derivation of each model and limit
ourselves to presenting here just the functional form ${\cal F}(x,\kappa^2)$
of the UGD as we implemented it in the numerical analysis.

%--------------------------------------------------------------
\subsection{An $x$-independent model (ABIPSW)}
%--------------------------------------------------------------
\label{Anikin}

The simplest UGD model is $x$-independent and merely coincides with
the proton impact factor~\cite{Anikin:2011sa}:
\begin{equation}
\label{ABIPSW}
{\cal F}(x,\kappa^2)= \frac{A}{(2\pi)^2\,M^2}
\left[\frac{\kappa^2}{M^2+\kappa^2}\right]\,,
\end{equation}
where $M$ corresponds to the non-perturbative hadronic scale, fixed as~$M = 1$~GeV. The constant $A$ is irrelevant when we consider the ratio $T_{11}/T_{00}$ for the $\rho$-meson leptoproduction, but it becomes essential to calculate cross sections. Therefore, we 
determined it by a global fit to all available data for polarized
cross sections, getting $A = 148.14 \text{ GeV}^2$, with a relative uncertainty below $0.05\%$.

%--------------------------------------------------
\subsection{Gluon momentum derivative}
%--------------------------------------------------

This UGD is given by
\begin{equation}
\label{xgluon}
{\cal F}(x, \kappa^2) = \frac{dxg(x, \kappa^2)}{d\ln \kappa^2}
\end{equation}
and encompasses the collinear gluon density $g(x, \mu_F^2)$, taken at
$\mu_F^2=\kappa^2$. It is based on the obvious requirement that, when
integrated over $\kappa^2$ up to some factorization scale, the UGD must
give the collinear gluon density. We have employed the {\tt CT14}
parametrization~\cite{Dulat:2015mca}, using the appropriate cutoff
$\kappa_{\text{min}} = 0.3$~GeV (see Section III A of Ref.~\cite{Bolognino:2018rhb} for further details).

%-------------------------------------------------------------------------------
\subsection{Ivanov--Nikolaev (IN) UGD: a soft-hard model}
%-------------------------------------------------------------------------------

The UGD proposed in Ref.~\cite{Ivanov:2000cm} is developed with the purpose
of probing different regions of the transverse momentum. In the large-$\kappa$
region, DGLAP parametrizations for $g(x, \kappa^2)$ are employed. Moreover,
for the extrapolation of the hard gluon densities to small $\kappa^2$, an
Ansatz is made~\cite{Nikolaev:1994cd}, which describes the color gauge
invariance constraints on the radiation of soft gluons by color singlet
targets. The gluon density at small $\kappa^2$ is supplemented by a
non-perturbative soft component, according to the color-dipole
phenomenology.

This model of UGD has the following two-component form:
\begin{equation}
{\cal F}(x,\kappa^2)= {\cal F}^{(B)}_\text{soft}(x,\kappa^2) 
{\kappa_{s}^2 \over 
	\kappa^2 +\kappa_{s}^2} + {\cal F}_\text{hard}(x,\kappa^2) 
{\kappa^2 \over 
	\kappa^2 +\kappa_{h}^2}\,,
\label{eq:4.7}
\end{equation}
where $\kappa_{s}^2 = 3$ GeV$^2$ and $\kappa_{h}^2 = [1 + 0.047\log^2(1/x)]^\frac{1}{2}$ GeV$^2$.

The soft term reads
\begin{equation}
\label{softterm}
{\cal F}^{(B)}_\text{soft}(x,\kappa^2) = a_\text{soft}
C_{F} N_{c} {\alpha_{s}(\kappa^2) \over \pi} \left( {\kappa^2 \over 
	\kappa^2 +\mu_\text{soft}^{2}}\right)^2 V_{\text N}(\kappa)\,,
\end{equation}
with $\mu_\text{soft} = 0.1$ GeV. The
parameter $a_\text{soft} = 2$ gives a measure of how important is the soft part
compared to the hard one. On the other hand, the hard component reads
\begin{equation}
\label{hardterm}
{\cal F}_\text{hard}(x,\kappa^2)= 
{\cal F}^{(B)}_{\text{pt}}(\kappa^2){{\cal F}_{\text{pt}}(x,Q_{c}^{2})
	\over {\cal F}_{\text{pt}}^{(B)}(Q_{c}^{2})}
\theta(Q_{c}^{2}-\kappa^{2}) +{\cal F}_{\text{pt}}(x,\kappa^2)
\theta(\kappa^{2}-Q_{c}^{2})\,,
\end{equation}	
where ${\cal F}_{\text{pt}}(x, \kappa^2)$ is related to the collinear gluon parton distribution function (PDF) as in Eq.~\eqref{xgluon} and $Q_{c}^2 = 3.26$ GeV$^2$ %is the soft-hard interface 
(see Section III A of Ref.~\cite{Bolognino:2018rhb} for further details).
We refer to Ref.~\cite{Ivanov:2000cm} for the expressions of the vertex
function $V_{\text N}(\kappa)$ and of ${\cal F}^{(B)}_{\text{pt}}(\kappa^2)$.
%is expressed through $F_2(\kappa, -\kappa)$,
%which assumes the meaning of the two-quark form factor of the nucleon, related
%to the single-quark form factor,
%\begin{equation}
%F_2(\kappa, -\kappa) = F_1\left(\frac{2N_{c}}{N_{c} -1}\kappa^2\right)\,.
%\end{equation}
Another relevant feature of this model is given by the choice of the coupling
constant. In this regard, the infrared freezing of strong coupling at leading
order (LO) is imposed by fixing $\Lambda_\text{QCD} = 200$ MeV:
\begin{equation}
\label{frozen}
\alpha_s(\mu^2) = \text{min} \left\{0.82, \, \frac{4 \pi}{\beta_0
	\log \left(\frac{\mu^2}{\Lambda^2_\text{QCD}}\right)}\right\}.
\end{equation}
We wish to stress that this model was successfully tested in 
the {\em unpolarized} electroproduction of vector mesons at HERA.

%-----------------------------------------------------------------------------
\subsection{Hentschinski--Sabio Vera--Salas (HSS) model}
%-----------------------------------------------------------------------------

This model, originally used in the study of DIS structure
functions~\cite{Hentschinski:2012kr}, takes the form of a convolution between
the BFKL gluon Green's function and a LO proton impact factor. It has been
employed in the description of single-bottom quark production at the LHC~\cite{Chachamis:2015ona}, to investigate the photoproduction of
$J/\Psi$ and $\Upsilon$ in~\cite{Bautista:2016xnp,Garcia:2019tne,Hentschinski:2020yfm} and to study the forward Drell--Yan invariant-mass distribution~\cite{Celiberto:2018muu}. We implemented
the formula given in Ref.~\cite{Chachamis:2015ona} (up to a $\kappa^2$ overall
factor needed to match our definition), which reads
\begin{equation}
\label{HentsUGD}
{\cal F}(x, \kappa^2, M_h) = \int_{-\infty}^{\infty}
\frac{d\nu}{2\pi^2}\ {\cal C} \  \frac{\Gamma(\delta - i\nu -\frac{1}{2})}
{\Gamma(\delta)}\ \left(\frac{1}{x}\right)^{\chi\left(\frac{1}{2}+i\nu\right)}
\left(\frac{\kappa^2}{Q^2_0}\right)^{\frac{1}{2}+i\nu}
\end{equation}
\[
\times \left\{ 1 +\frac{\bar{\alpha}^2_s \beta_0 \chi_0\left(\frac{1}{2}
	+i\nu\right)}{8 N_c}\log\left(\frac{1}{x}\right)
\left[-\psi\left(\delta-\frac{1}{2} - i\nu\right)
-\log\frac{\kappa^2}{M_h^2}\right]\right\}\,,
\]
where $\beta_0=\frac{11 N_c-2 N_f}{3}$, with $N_f$ the number of
active quarks (put equal to four in the following),
$\bar{\alpha}_s = \dfrac{\alpha_s\left(\mu^2\right) N_c}{\pi}$,
with $\mu^2 = Q_0 M_h$, and $\chi_0(\frac{1}{2} + i\nu)\equiv \chi_0(\gamma)
= 2\psi(1) - \psi(\gamma) - \psi(1-\gamma)$ is  (up to the factor
$\bar\alpha_s$) the LO eigenvalue of the BFKL kernel, 
with $\psi(\gamma)$ the logarithmic derivative of the Euler Gamma
function. 
Here, $M_h$ plays the role of the hard scale which in our case can be identified
with the photon virtuality, $\sqrt{Q^2}$.
In Eq.~\eqref{HentsUGD}, $\chi(\gamma)$ (with $\gamma = \frac{1}{2} + i\nu$)
is the NLO eigenvalue of the BFKL kernel, collinearly improved and BLM
optimized. It reads
\begin{equation}
\chi(\gamma) = \bar{\alpha}_s\chi_0(\gamma)+\bar{\alpha}^2_s\chi_1(\gamma)
-\frac{1}{2}\bar{\alpha}^2_s\chi^\prime_0(\gamma)\,\chi_0(\gamma)
+ \chi_{RG}(\bar{\alpha}_s, \gamma)\,,
\end{equation}
with $\chi_1(\gamma)$ and $\chi_{RG}(\bar{\alpha}_s, \gamma)$ given in
Section~2 of Ref.~\cite{Chachamis:2015ona}.

This UGD model is characterized by a peculiar parametrization for the proton
impact factor, whose expression is
\begin{equation}
\Phi_p(q, Q^2_0) = \frac{{\cal C}}{2\pi \Gamma(\delta)}
\left(\frac{q^2}{Q^2_0}\right)^\delta e^{-\frac{q^2}{Q^2_0}}\,,
\end{equation}
which depends on three parameters $Q_0$, $\delta$ and ${\cal C}$ which
were fitted to the combined HERA data for the $F_2(x)$ proton structure
function. We adopted here the so-called
{\em kinematically improved} values (see Section III A of Ref.~\cite{Bolognino:2018rhb} for further
details) and given by
\begin{equation}
\label{ki}
Q_0 = 0.28\,\text{GeV}, \qquad \delta = 6.5, \qquad {\cal C} = 2.35 \;.
\end{equation}

%--------------------------------------------------------------------
\subsection{Golec-Biernat--W{\"u}sthoff (GBW) UGD}
%--------------------------------------------------------------------

This UGD parametrization derives from the effective dipole cross section
$\hat{\sigma}(x,r)$ for the scattering of a $q\bar{q}$ pair off a
nucleon~\cite{GolecBiernat:1998js},
\begin{equation}
\hat{\sigma}(x, r^2) = \sigma_0 \left\{1-\exp\left(-\frac{r^2}{4R^2_0(x)}
\right)\right\}\,,
\end{equation}
through a reverse Fourier transform of the expression 
\begin{equation}
\sigma_0 \left\{1-\exp\left(-\frac{r^2}{4R^2_0(x)}
\right)\right\}={2 \pi \over N_c} \int \frac{d^2\kappa}{\kappa^4} \alpha_s {\cal F}(x,\kappa^2)
\left(1-\exp(i \vec{\kappa}\cdot\vec{r})\right)\left(1-\exp(-i \vec{\kappa}
\cdot\vec{r})\right)\,,
\end{equation}
\begin{equation}
\alpha_s {\cal F}(x,\kappa^2)= N_c \kappa^4 \sigma_0 \frac{R^2_0(x)}{4 \pi^2}
e^{-\kappa^2 R^2_0(x)}\,,
\end{equation}
with 
\begin{equation}
R^2_0(x) = \frac{1}{{\text{GeV}}^2} \left(\frac{x}{x_0}\right)^{\lambda_p}
\end{equation}
and  the following values
\begin{equation}
\sigma_0 = 23.03\,\text{mb}, \qquad \lambda_p = 0.288, \qquad x_0 = 3.04 \cdot 10^{-4}\,.
\end{equation}
The normalization $\sigma_0$ and the parameters $x_0$ and $\lambda_p > 0$ of
$R^2_0(x)$ have been determined by a global fit to $F_2(x)$ in the
region $x < 0.01$.

%-------------------------------------------------------------
\subsection{Watt--Martin--Ryskin (WMR) model}
%-------------------------------------------------------------

This model is based on the idea that the $\kappa$ dependence of the UGD 
comes from the last step of the evolution ladder.

The UGD introduced in Ref.~\cite{Watt:2003mx} reads
\[
{\cal F}(x,\kappa^2,\mu^2) = T_g(\kappa^2,\mu^2)\,\frac{\alpha_s(\kappa^2)}
{2\pi}\,\int_x^1\!dz\;\left[\sum_q P_{gq}(z)\,\frac{x}{z}q\left(\frac{x}{z},
\kappa^2\right) + \right.\nonumber
\]
\begin{equation}
\label{WMR_UGD}
\left. \hspace{6.5cm} P_{gg}(z)\,\frac{x}{z}g\left(\frac{x}{z},\kappa^2\right)\,\Theta\left(\frac{\mu}{\mu+\kappa}-z\right)\,\right]\,,
\end{equation}
where the term
\begin{equation}
\label{WMR_Tg}
T_g(\kappa^2,\mu^2) \!= \exp\!\left(-\int_{\kappa^2}^{\mu^2}\!d\kappa_t^2\,
\frac{\alpha_s(\kappa_t^2)}{2\pi}\,\!\left( \int_{z^\prime_{{\text{min}}}}^{z^\prime_{{\text{max}}}}
\!dz^\prime\;z^\prime \,P_{gg}(z^\prime ) + N_f\!\!\int_0^1\!dz^\prime\,P_{qg}(z^\prime)
\right)\right)\,\!,
\end{equation}
gives the probability of evolving from the scale $\kappa$ to the
scale $\mu$ without parton emission. Here $z^\prime_{\mathrm{max}}\equiv
1-z^\prime_{\mathrm{min}}=\mu/(\mu+\kappa_t)$; $N_f$ is the number of active quarks.
This UGD model depends on an extra-scale $\mu$, which we fixed at $Q$.
The splitting functions $P_{qg}(z)$ and $P_{gg}(z)$ are given by
\[
P_{qg}(z) = T_R\,[z^2 + (1-z)^2]\;,
\]
\[
P_{gg}(z) = 2\,C_A \left[\dfrac{1}{(1-z)_+} + \dfrac{1}{z}- 2 +z(1-z)\right]
+ \left(\frac{11}{6}C_A - \frac{N_f}{3}\right) \delta(1 -z)\;,
\]
with the plus-prescription defined as
\begin{equation}
\label{pluspre}
\int_{a}^{1} dz \frac{F(z)}{(1-z)_+} = \int_{a}^{1} dz \frac{F(z) - F(1)}{(1-z)}
- \int_{0}^{a} dz \frac{F(1)}{(1 -z)}\,.
\end{equation}

%-----------------------------------------------------------------------------------------------
\subsection{Bacchetta--Celiberto--Radici--Taels (BCRT) distribution}
%------------------------------------------------------------------------------------------------

We include in our analysis the unpolarized distribution part of the set of leading-twist $T$-even transverse-momentum-dependent (TMD) gluon PDFs calculated in Ref.\tcite{Bacchetta:2020vty} (see also Ref.\tcite{Celiberto:2021zww}), suited for studies in wider kinematic ranges at new-generation collider machines, such as the EIC\tcite{AbdulKhalek:2021gbh}, HL-LHC\tcite{Chapon:2020heu}, and NICA-SPD\tcite{Arbuzov:2020cqg}. One has
\begin{equation}
 {\cal F}(x,\kappa^2) = \kappa^2 \int_M^{\infty} d M_X \, \rho_X (M_X) \, \hat{f}_1^g(x, \kappa^2; M_X) \;,
\label{BCRT}
\end{equation}
Here, $\hat{f}_1^g(x, \kappa^2; M_X)$ stands for the distribution of unpolarized gluons in an unpolarized proton, calculated in the so-called \emph{spectator-model} approximation, namely where the remainders of the proton after gluon emission are treated as a single spin-1/2 spectator particle of mass $M_X$
\begin{eqnarray}
\label{BCRT_f1}
 \hat{f}_1^g(x, \kappa^2; M_X) &= &\Big[ \big( 2 M x g_1(\kappa^2) - x (M+M_X) g_2(\kappa^2) \big)^2 \, \big[ (M_X - M (1-x) )^2 + \kappa^2 \big] \nonumber \\
 & &\quad + \, 2 \kappa^2 \, (\kappa^2 + x M_X^2) \, g_2(\kappa^2)^2 + 2 \kappa^2 M^2 \, (1-x) \, (4 g_1(\kappa^2)^2 - x g_2(\kappa^2)^2 ) \Big]  \nonumber \\ 
 & &\times \; \Big[ (2 \pi)^3 \, 4 x M^2 \, (x M_X^2 - x (1 - x) M^2 + \kappa^2)^2 \Big]^{-1} \; .
\end{eqnarray}
In Eq.~\eqref{BCRT_f1} $M$ is the proton mass, whereas $g_{1,2}$ couplings depict the effective proton-gluon-spectator vertex interaction
\begin{equation}
 g_{1,2}(\kappa^2) = G_{1,2} \, \frac{(1 - x) \left[ \kappa^2 + x M_X^2 - x (1 - x) M^2 \right]}{\left[ \kappa^2 + x M_X^2 - x (1 - x) M^2 + (1 - x) \Lambda_X^2 \right]^2} \;.
\label{BRCT_g12}
\end{equation}
The spectral function $\rho_X$ in Eq.~\eqref{BCRT} weighs $\hat{f}_1$ over $M_X$ in a continuous range and its expression reads
\begin{equation}
 \rho_X (M_X) = \mu^{2a} \left( \frac{A}{B + \mu^{2b}} + \frac{C}{\pi \sigma} e^{-\frac{(M_X - D)^2}{\sigma^2}} \right) \;,
\label{BCRT_rhoX}
\end{equation}
where $\mu^2 = M_X^2 -M^2$, and the model parameters in Eqs.~\eqref{BRCT_g12} and~\eqref{BCRT_rhoX} were obtained by a simultaneous fit of the $\kappa$-integral of spectator-model unpolarized and helicity functions on {\tt NNPDF3.1sx}\tcite{Ball:2017otu} and {\tt NNPDFpol1.1}\tcite{Nocera:2014gqa} collinear PDFs, respectively, and they encode effective small-$x$ effects coming from the BFKL resummation. In our study we employ values for these parameters obtained by averaging over the whole set of replicas provided (see Section~3 of Ref.\tcite{Bacchetta:2020vty} for details). They read: $G_1 = 1.51$ GeV$^2$, $G_2 = 0.414$ GeV$^2$, $\Lambda_X = 0.472$ GeV, $A = 6.1$, $a = 0.82$, $b = 1.43$, $C = 371$, $D = 0.548$ GeV, and $\sigma = 0.52$ GeV.

We stress that exploratory studies by using this TMD model as a UGD are
justified by the fact that we are considering observables for the
exclusive $\rho$-meson leptoproduction in the forward limit. In a more
general, off-forward case one should consider rather models for the gluon generalized parton distribution (GPD), instead of TMDs.

%--------------------
\section{Results}
%--------------------
\label{results}

All the results shown in this Section were obtained by making use of the \emph{Leptonic-Exclusive-Amplitudes} ({\tt LExA}) modular interface as implemented in the {\tt JETHAD} code\tcite{Celiberto:2020wpk}.

\begin{figure}
\centering

\includegraphics[scale=0.55,clip]{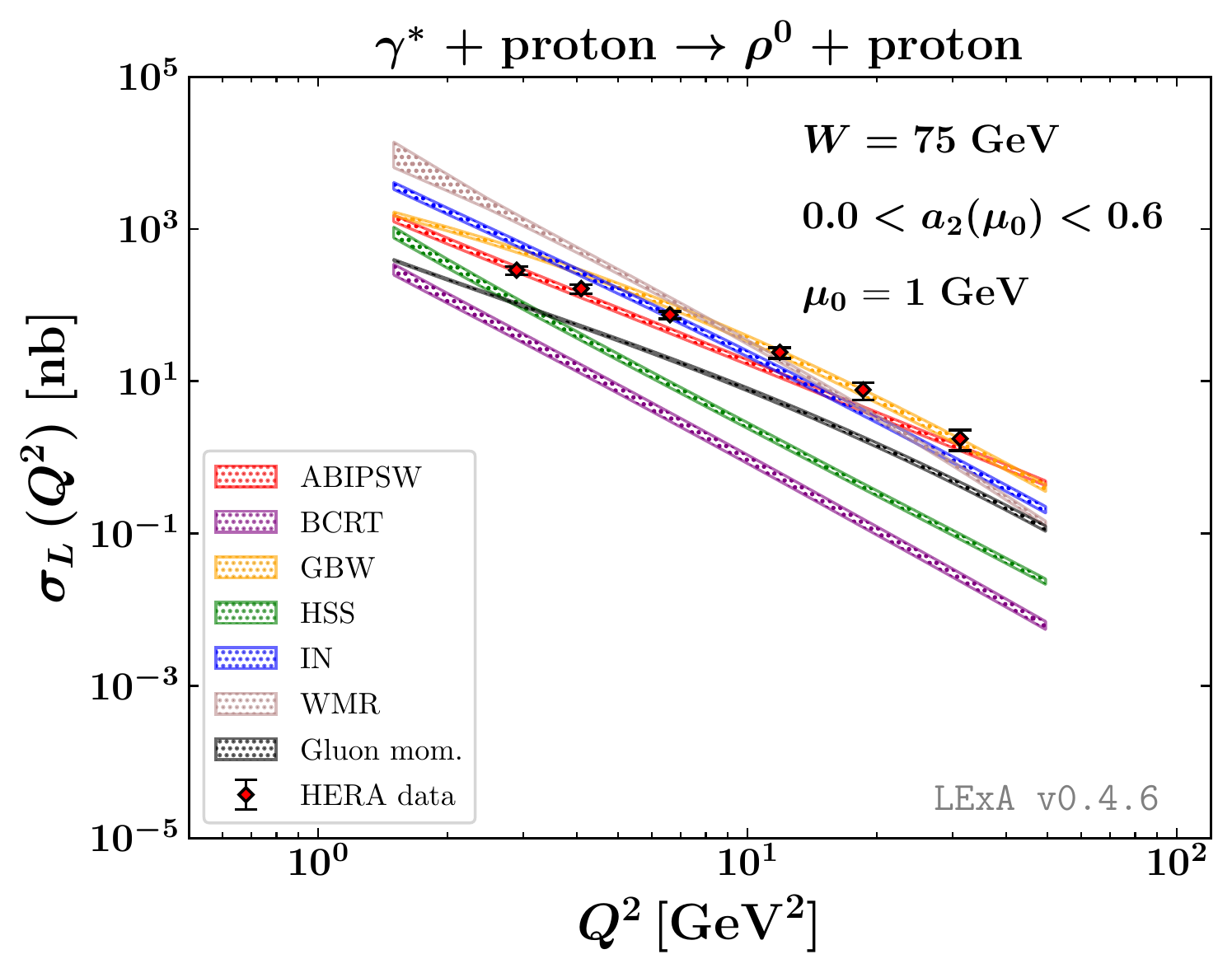}
\includegraphics[scale=0.55,clip]{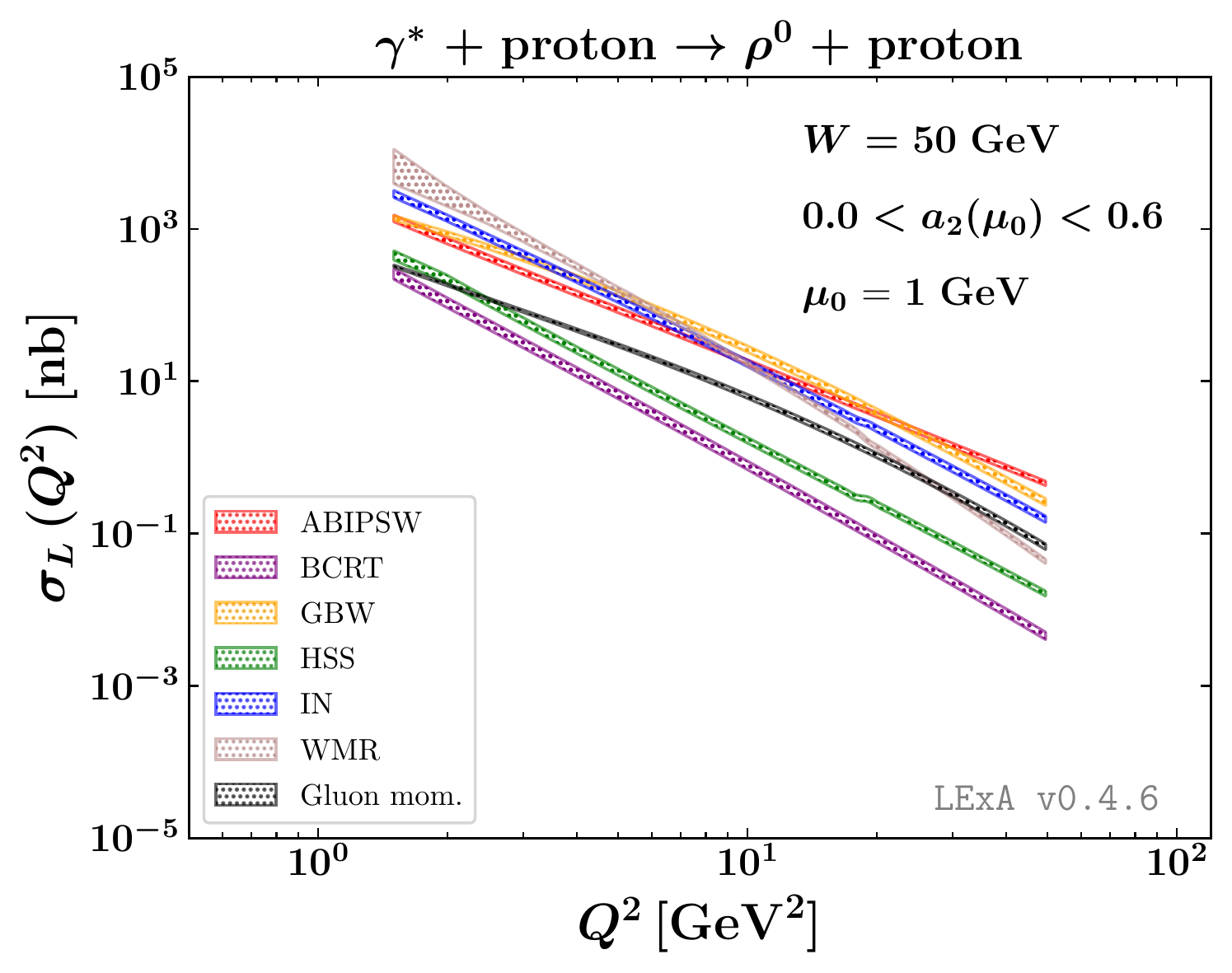}
\\
\includegraphics[scale=0.55,clip]{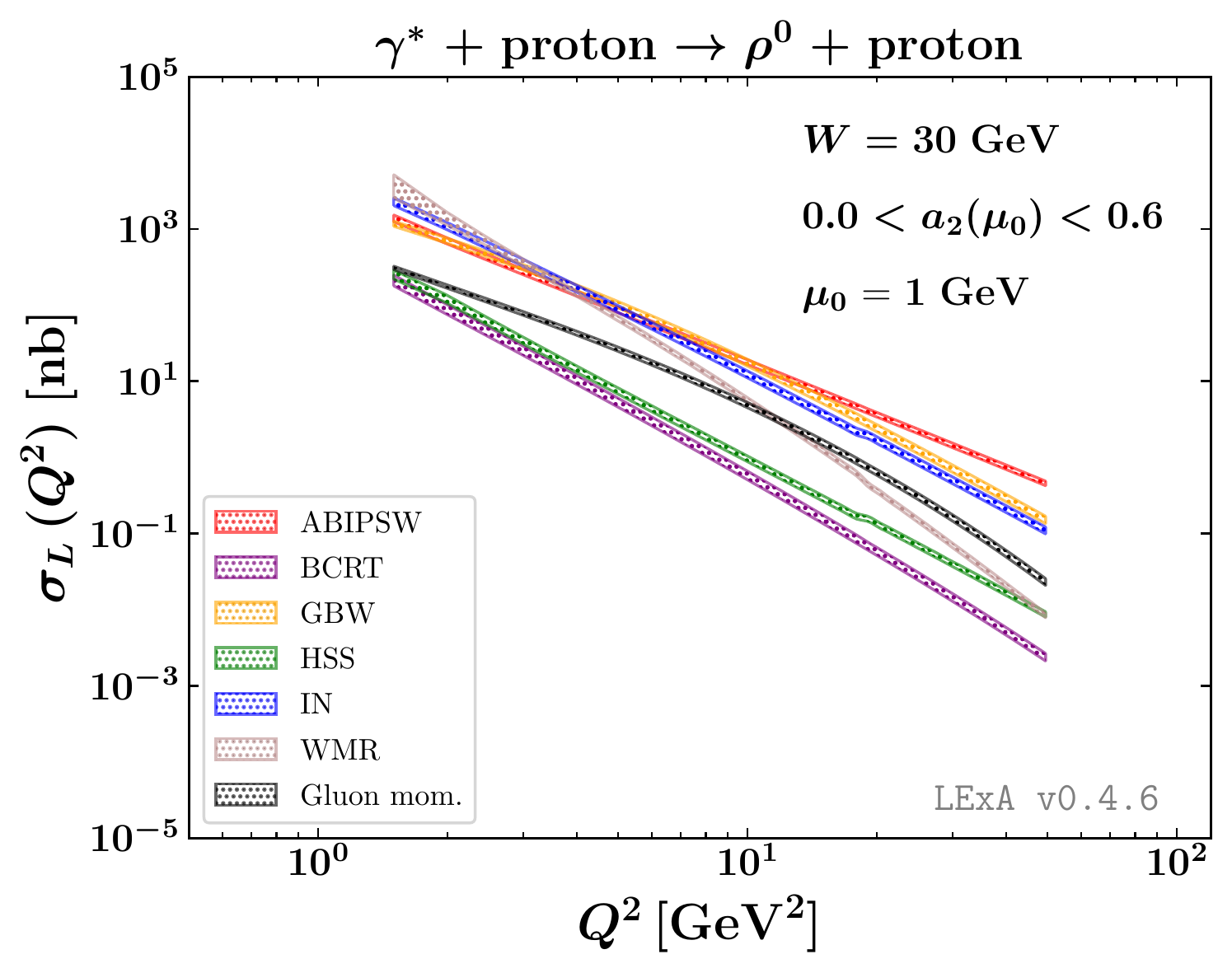}
\includegraphics[scale=0.55,clip]{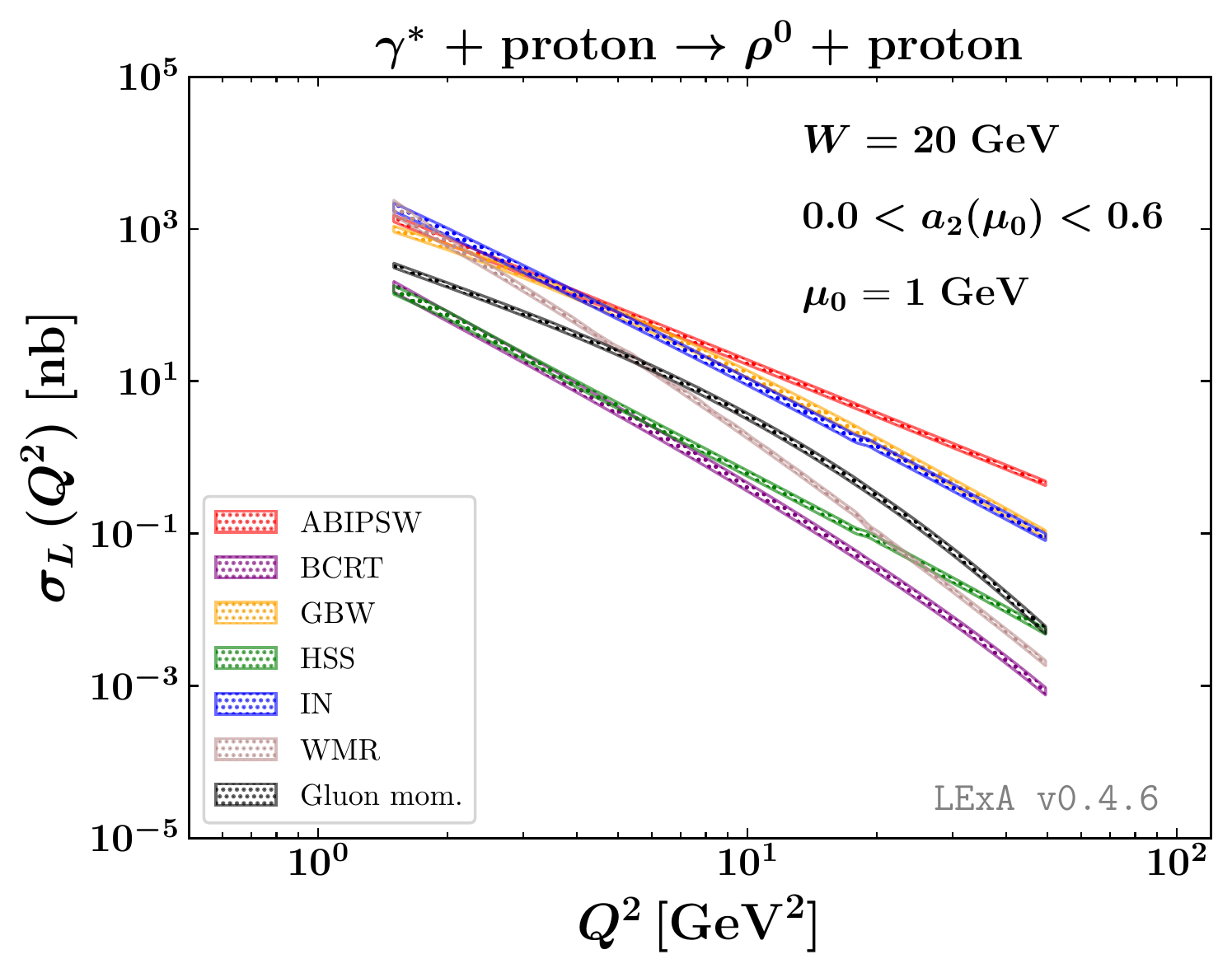}
\caption{$Q^2$-dependence of the longitudinally polarized cross section,
  $\sigma_L$, for all the considered UGD models, at $W = 75$ GeV
  together with the HERA data (left upper panel) and at $W = 20, 30, 50$
  GeV for EIC (the remaining panels). 
Uncertainty bands represent the effect of varying $a_2(\mu_0 = 1\,$\rm GeV$)$ between $0.0$ and $0.6$.}
\label{fig:sigma_L_all}
\end{figure}

\begin{figure}
\centering

\includegraphics[scale=0.55,clip]{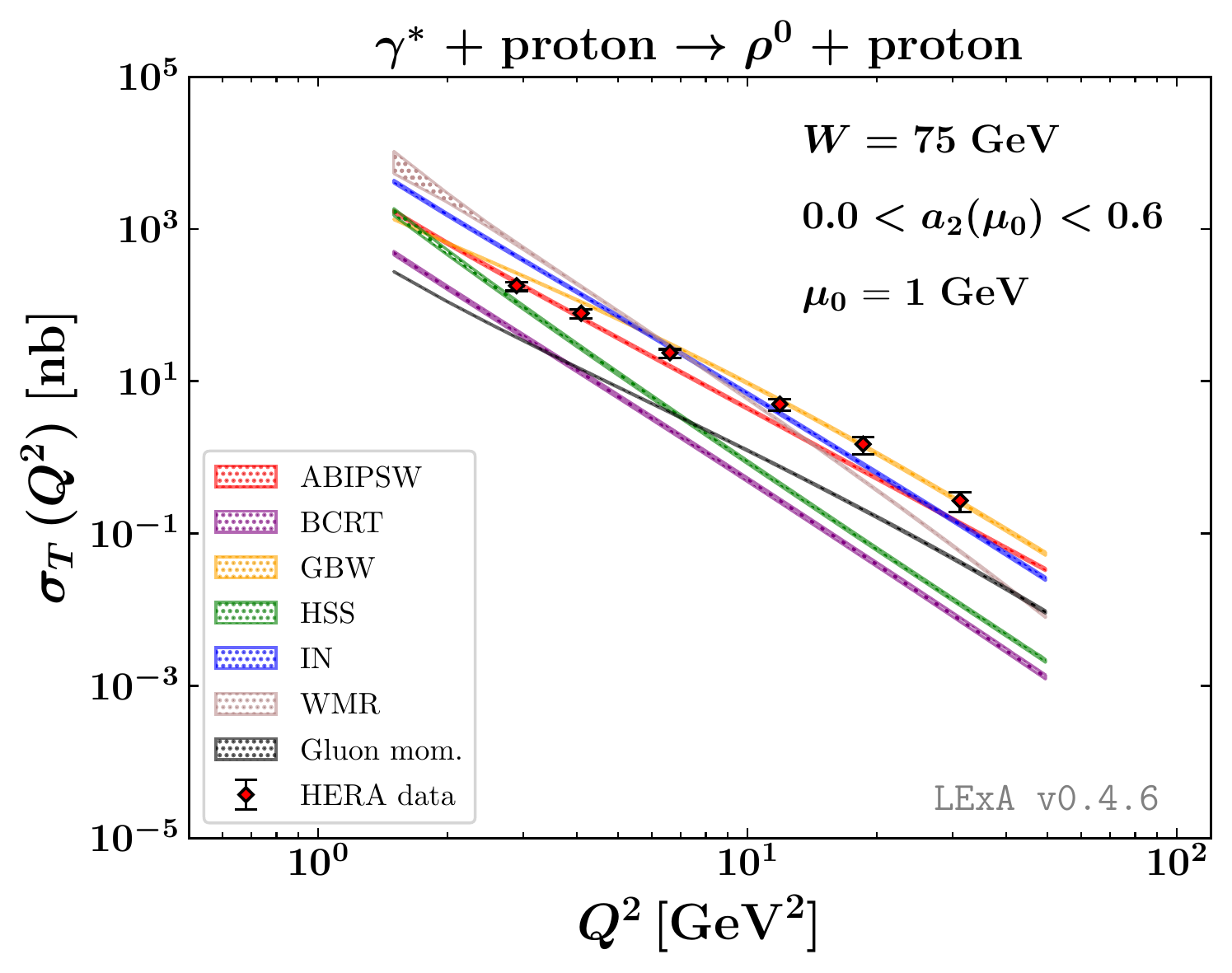}
\includegraphics[scale=0.55,clip]{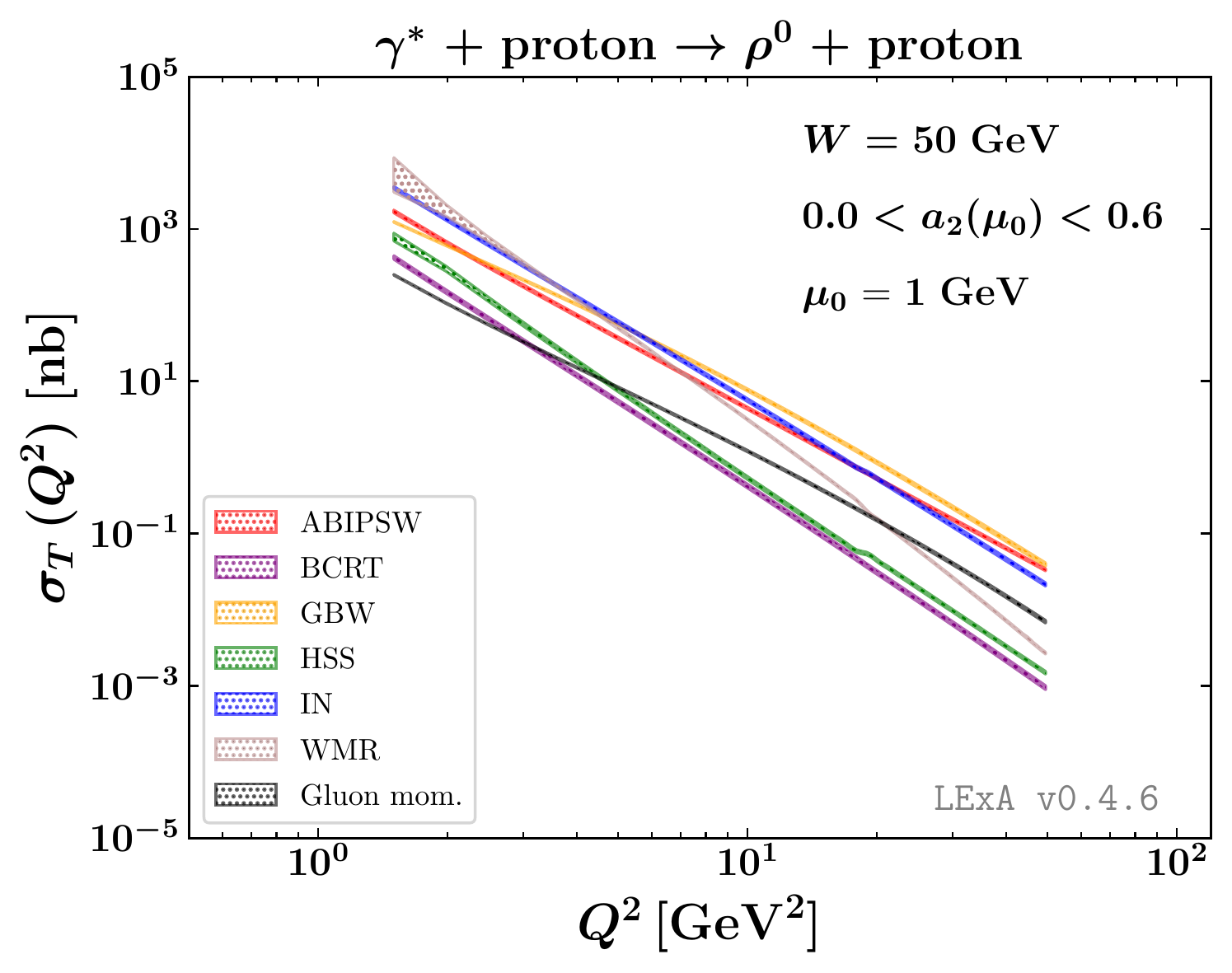}
\\
\includegraphics[scale=0.55,clip]{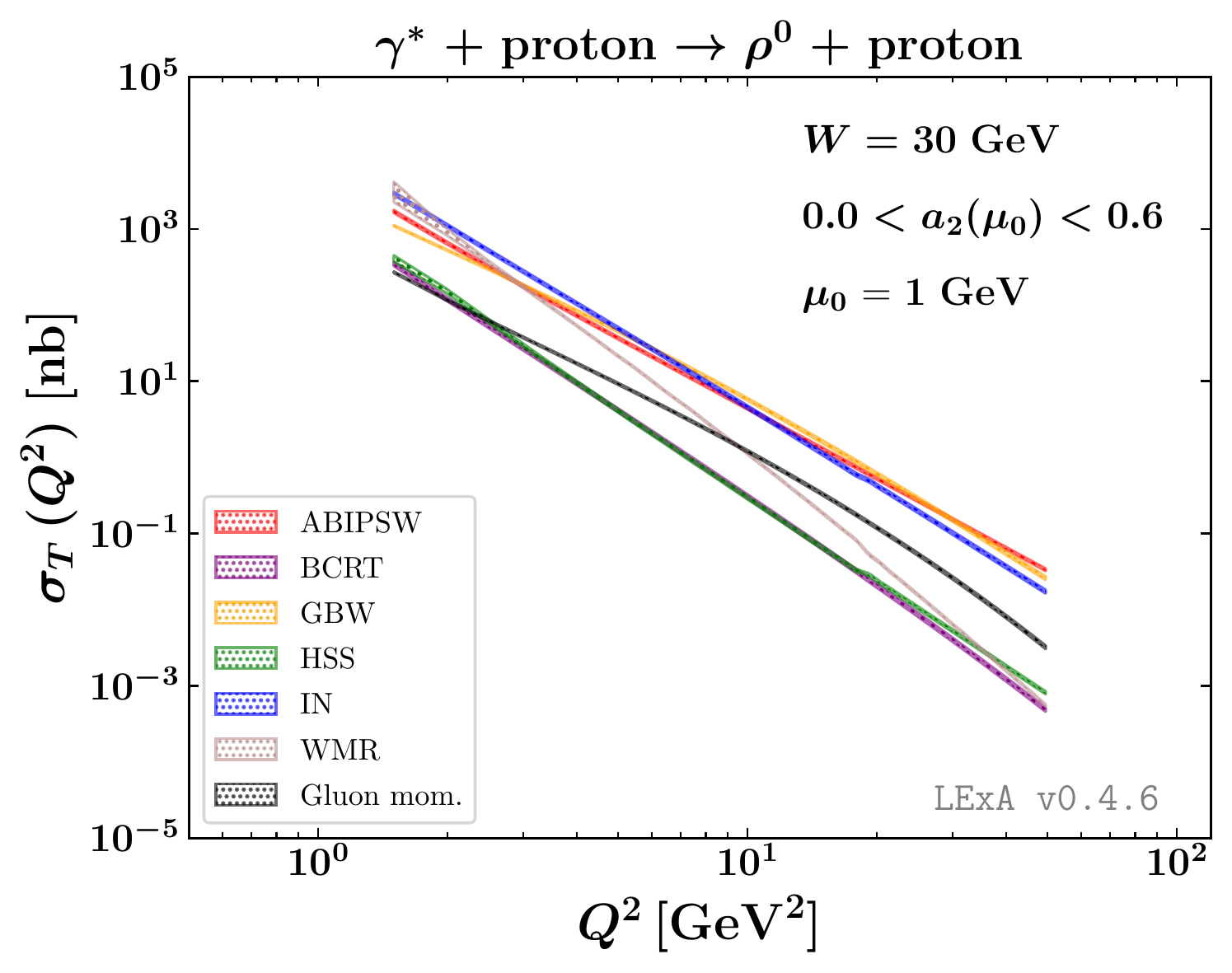}
\includegraphics[scale=0.55,clip]{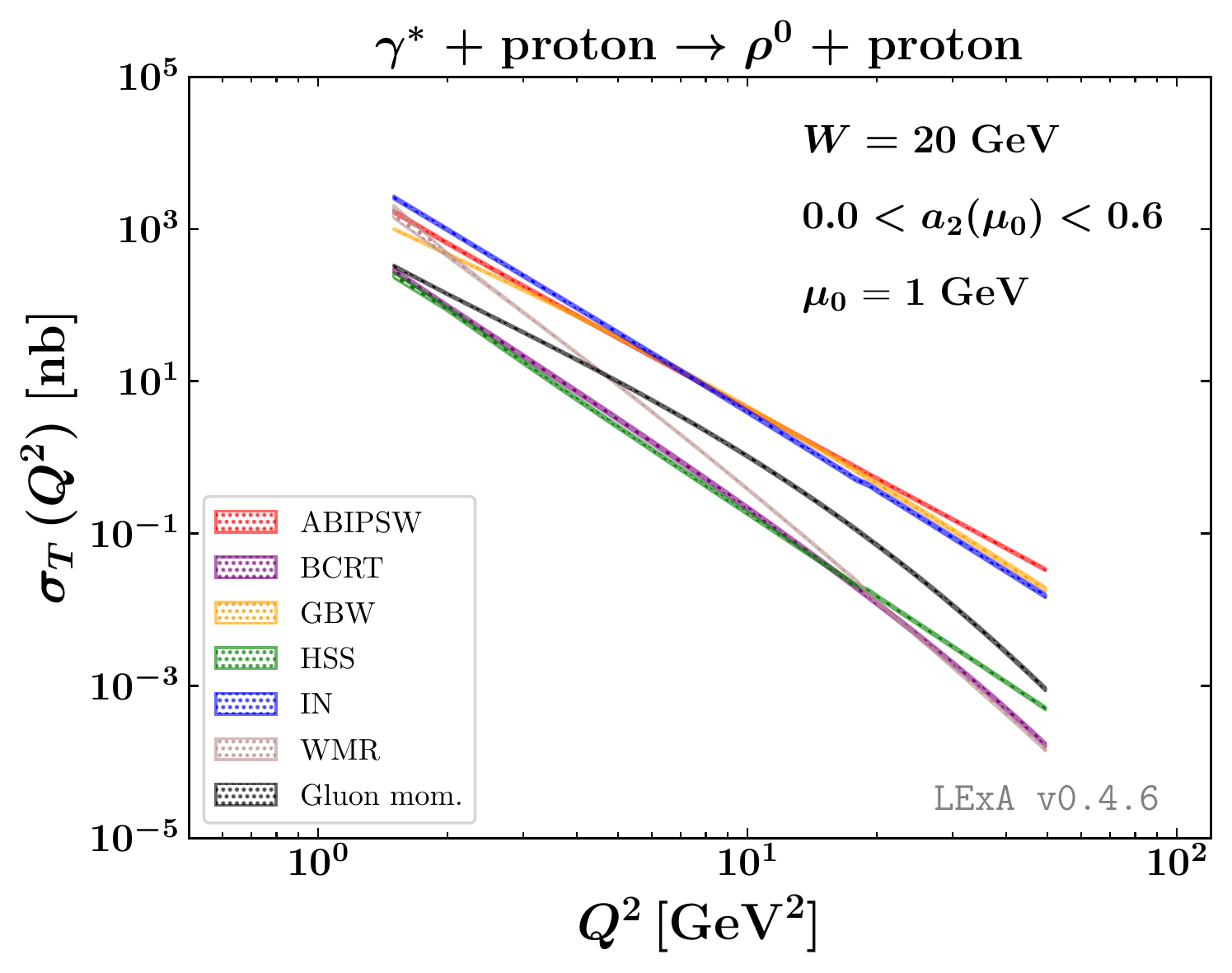}
\caption{$Q^2$-dependence of the transversely polarized cross section,
  $\sigma_T$, for all the considered UGD models, at $W = 75$ GeV
  together with the HERA data (left upper panel) and at $W = 20, 30, 50$
  GeV for EIC (the remaining panels). Uncertainty bands represent the effect
  of varying 
$a_2(\mu_0 = 1\,$\rm GeV$)$ between $0.0$ and $0.6$.}
\label{fig:sigma_T_all}
\end{figure}

\begin{figure}
\centering

\includegraphics[scale=0.55,clip]{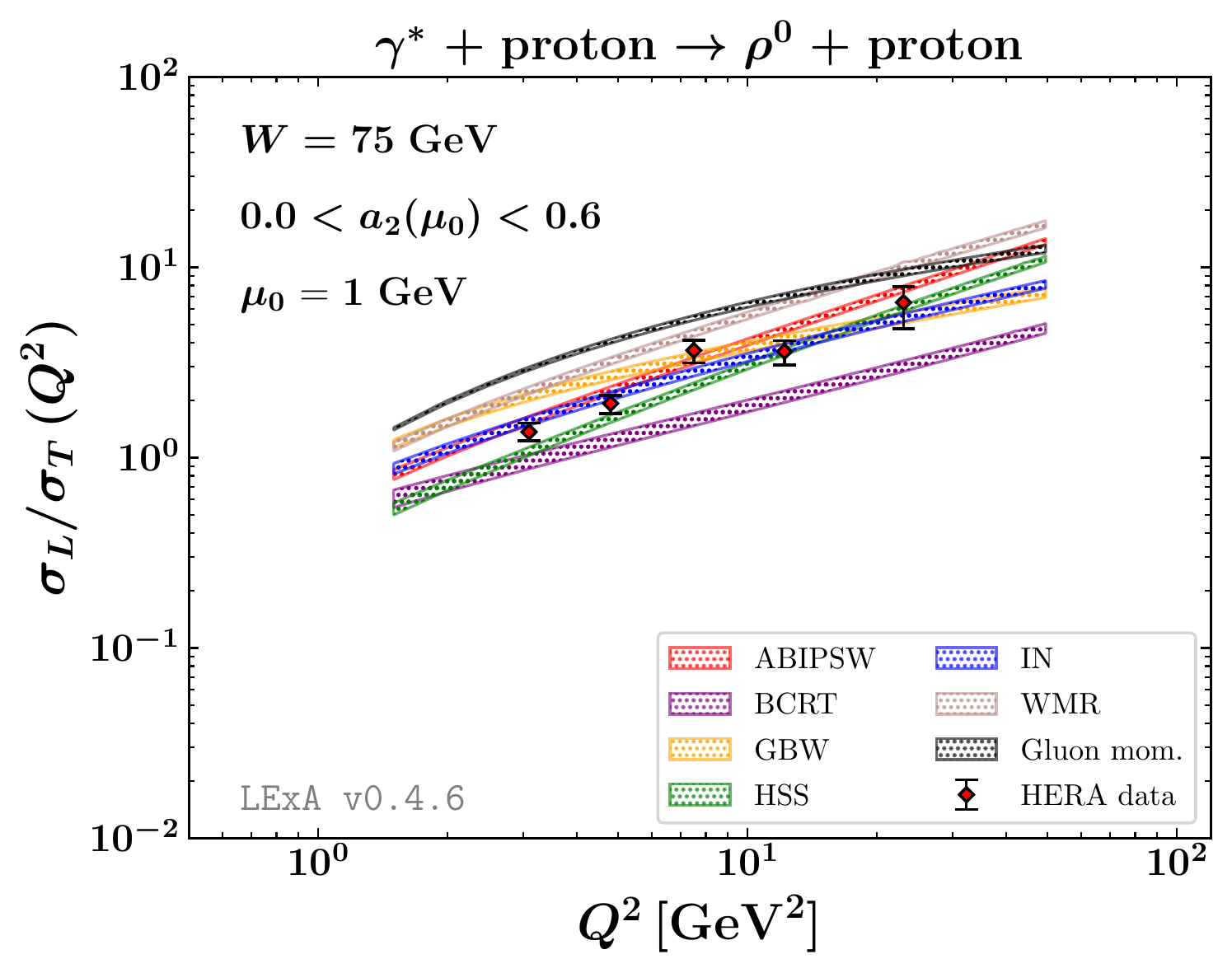}
\includegraphics[scale=0.55,clip]{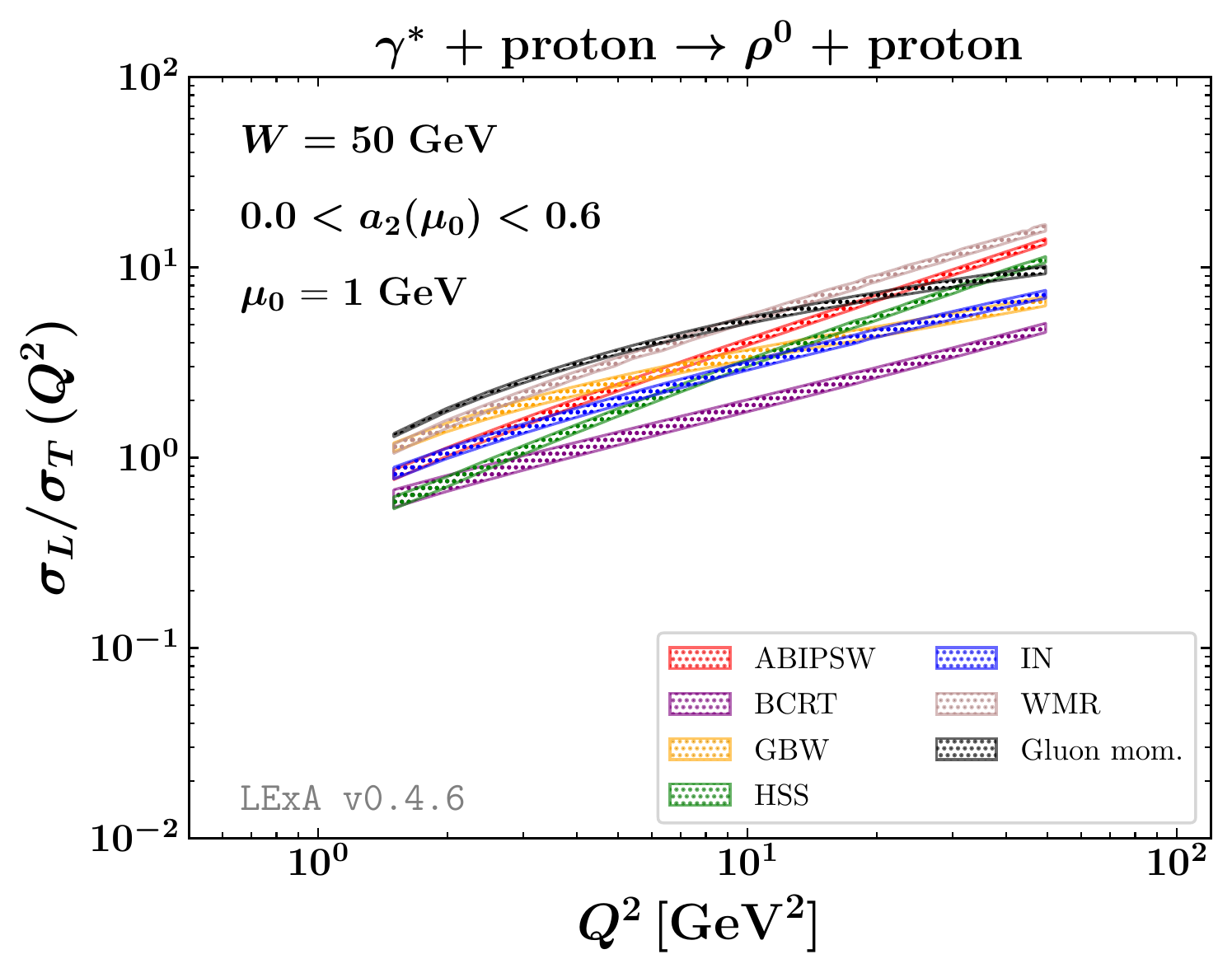}
\\
\includegraphics[scale=0.55,clip]{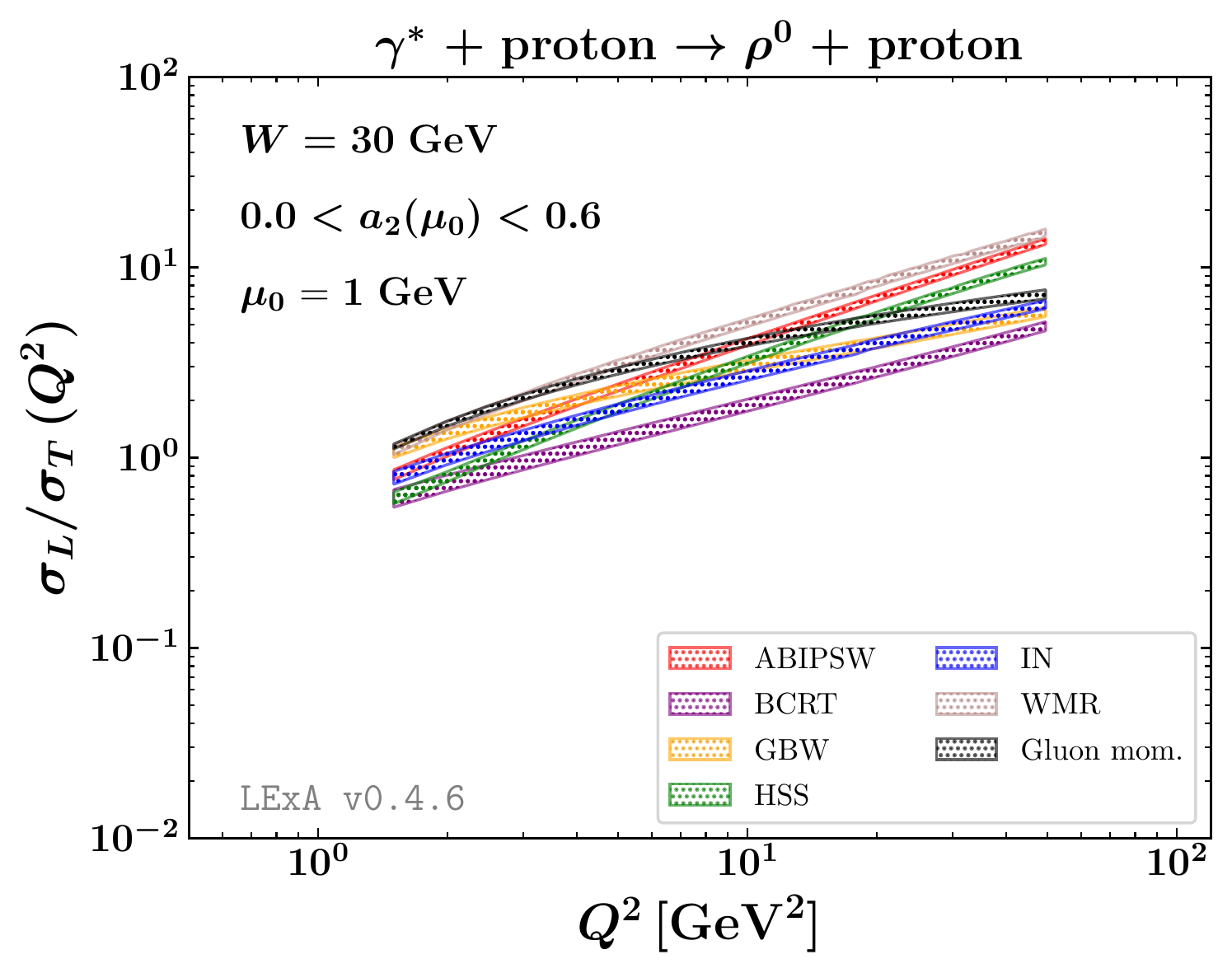}
\includegraphics[scale=0.55,clip]{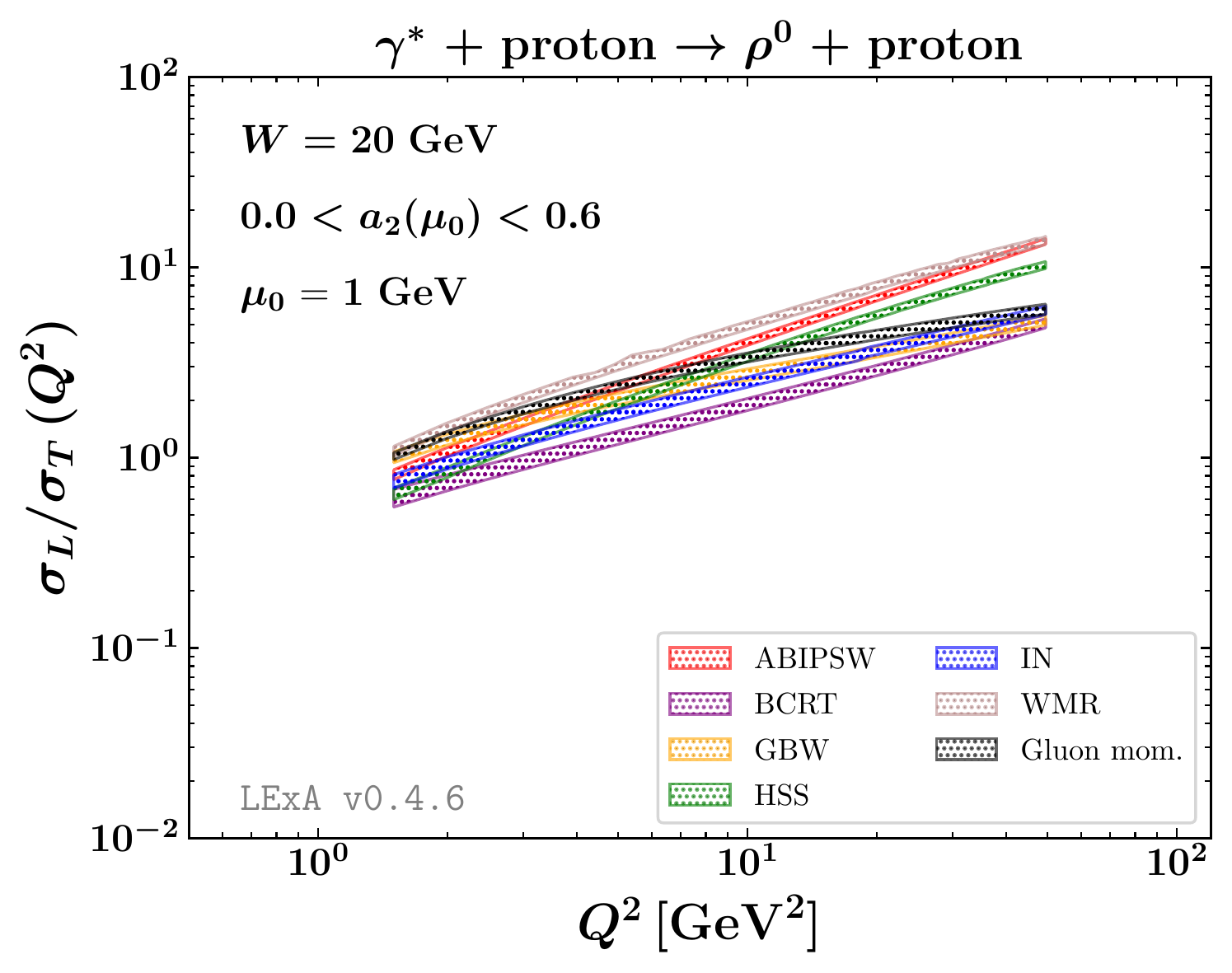}
\caption{$Q^2$-dependence of the polarized cross-section ratio,
  $\sigma_L/\sigma_T$, for all the considered UGD models, at $W = 75$
  GeV together with the HERA data (left upper panel) and at $W = 20, 30,
  50$ GeV for EIC (the remaining panels). Uncertainty bands represent the
  effect of varying $a_2(\mu_0 = 1\,$\rm GeV$)$ between $0.0$ and $0.6$.}
\label{fig:sigma_R_all}
\end{figure}

\begin{figure}
\centering

\includegraphics[scale=0.55,clip]{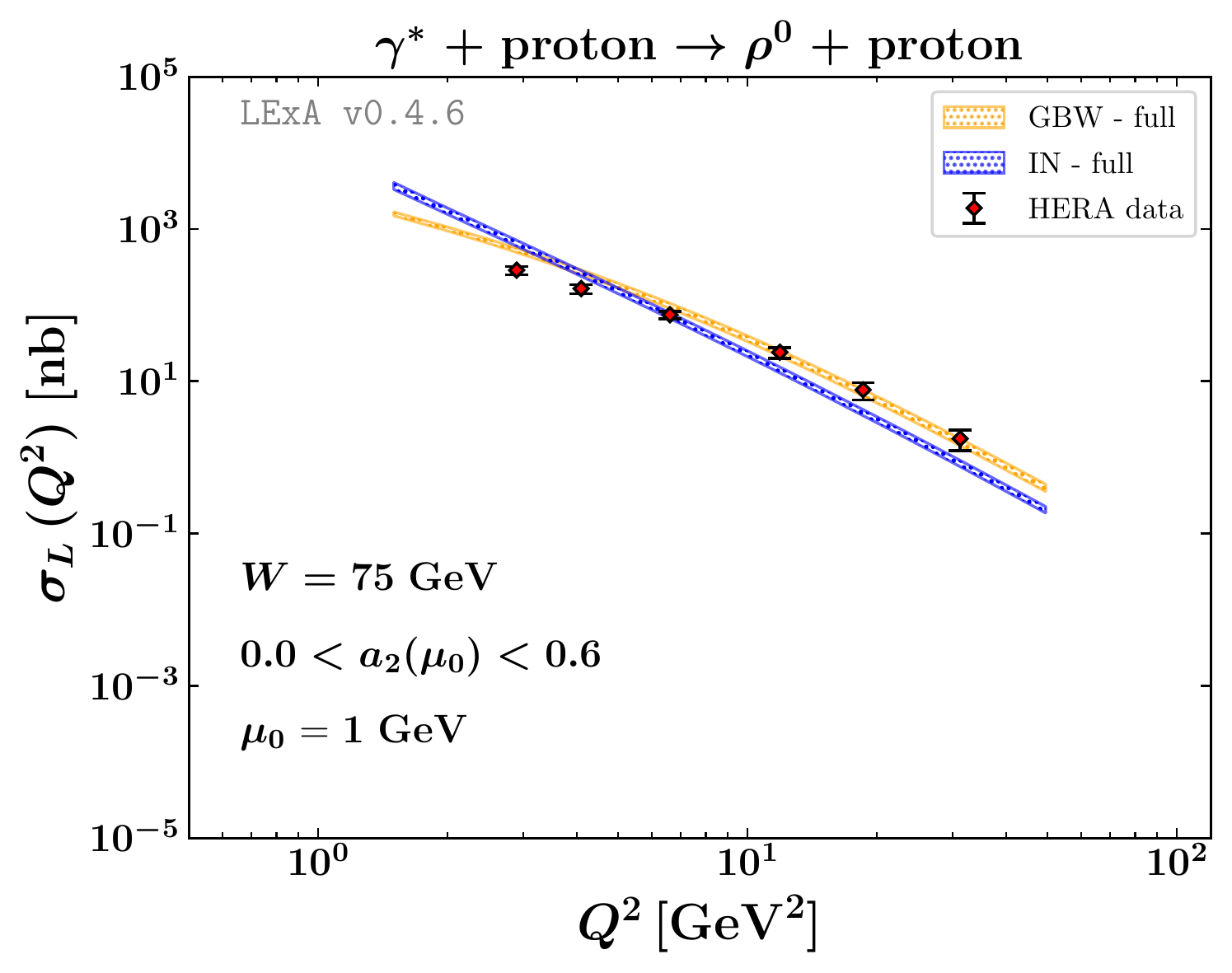}
\includegraphics[scale=0.55,clip]{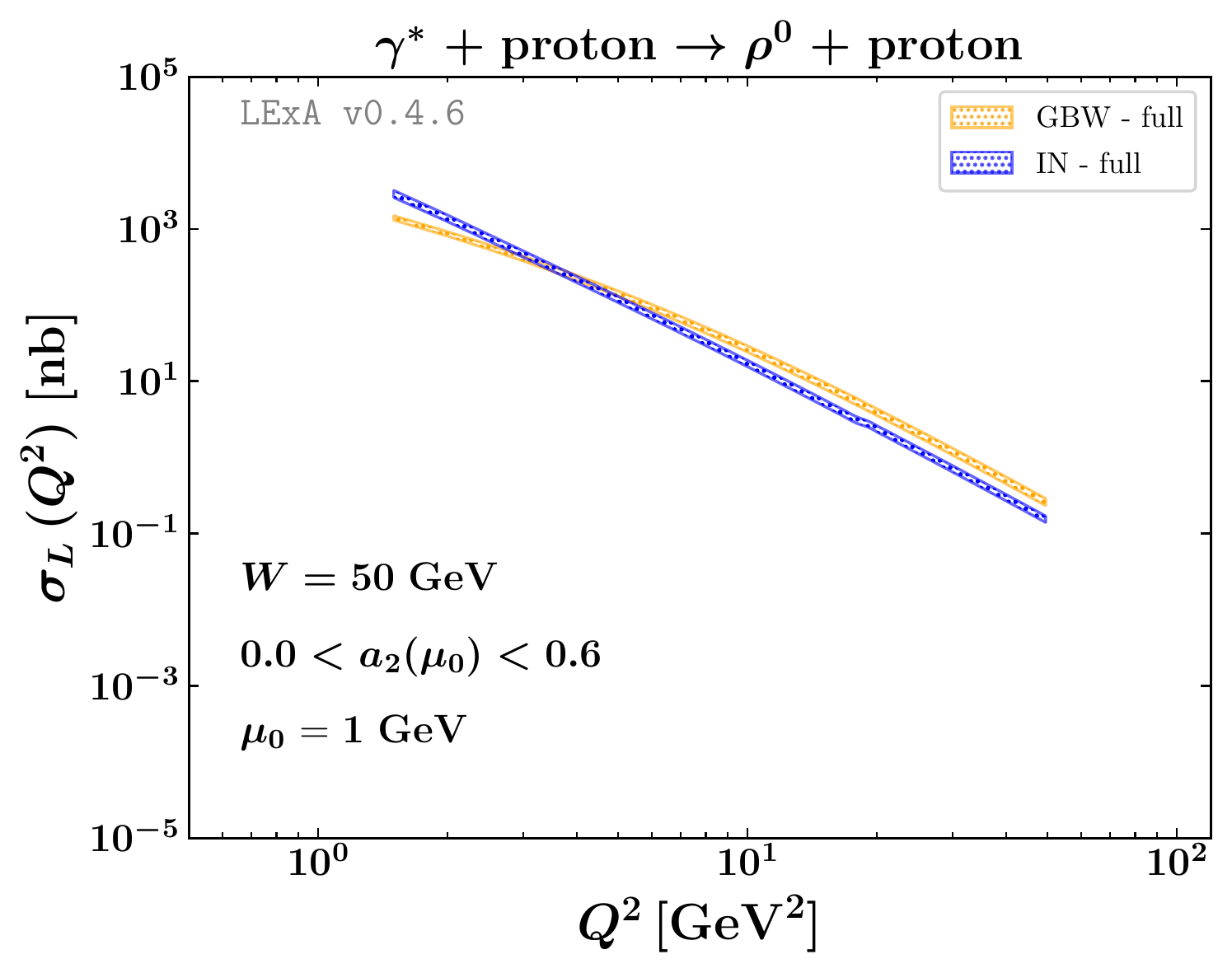}
\\
\includegraphics[scale=0.55,clip]{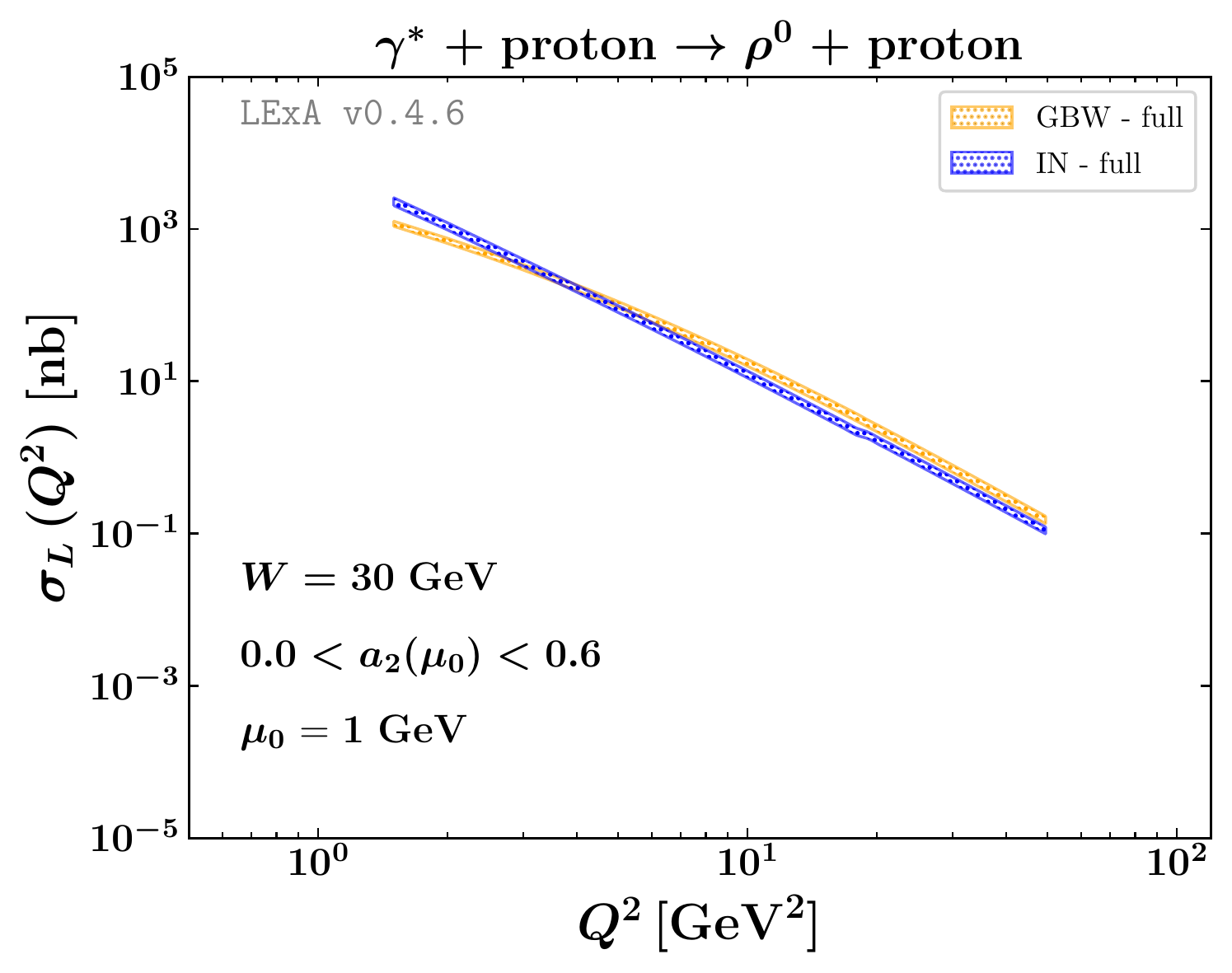}
\includegraphics[scale=0.55,clip]{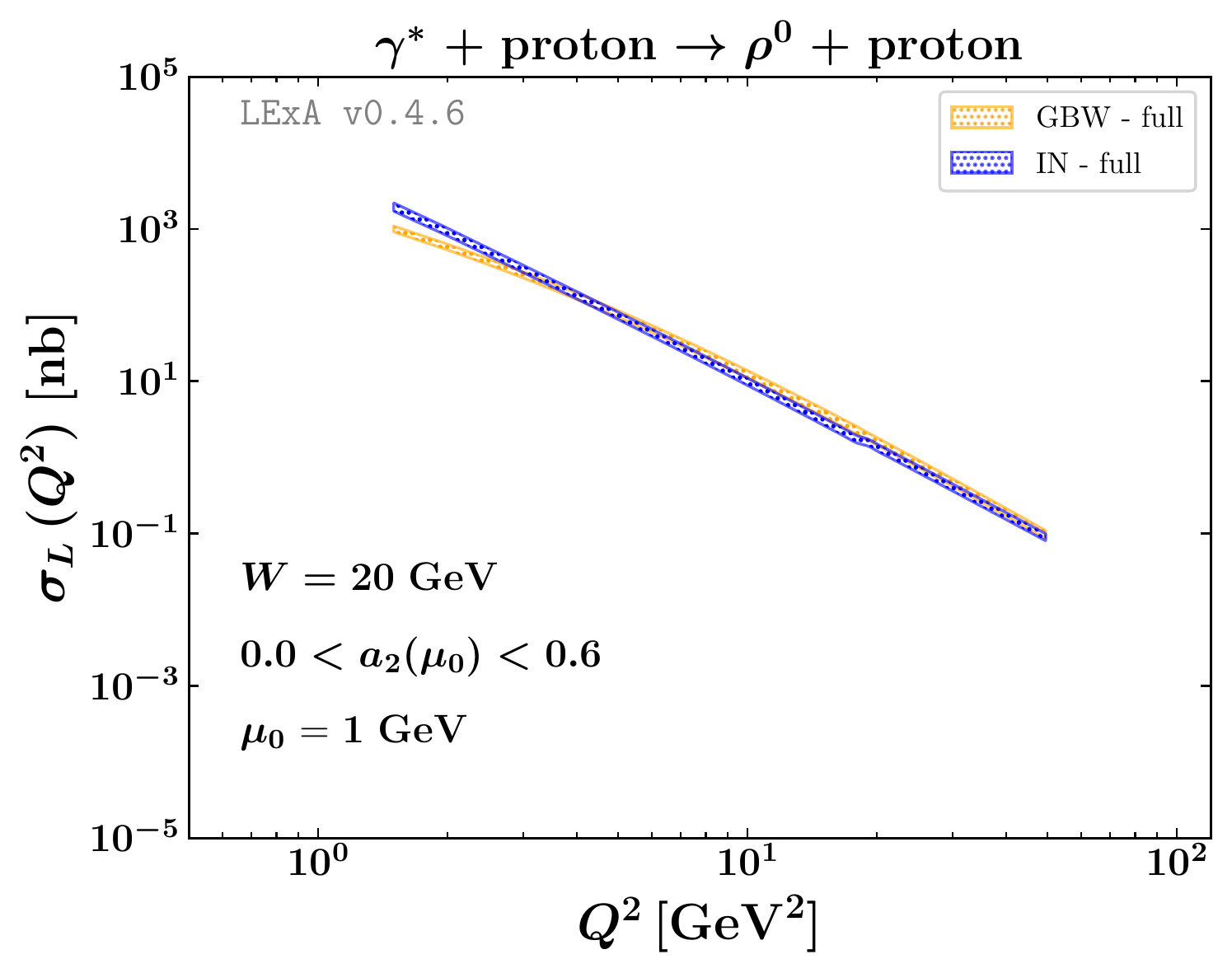}
\caption{$Q^2$-dependence of the longitudinally polarized cross section,
  $\sigma_L$, for GBW and IN UGD models, at $W = 75$ GeV together with
  the HERA data (left upper panel) and at $W = 20, 30, 50$ GeV EIC (the
  remaining panels). 
Uncertainty bands represent the effect of varying $a_2(\mu_0 = 1\,$\rm
GeV$)$ between $0.0$ and $0.6$.}
\label{fig:sigma_L_GBW}
\end{figure}

\begin{figure}
\centering

\includegraphics[scale=0.55,clip]{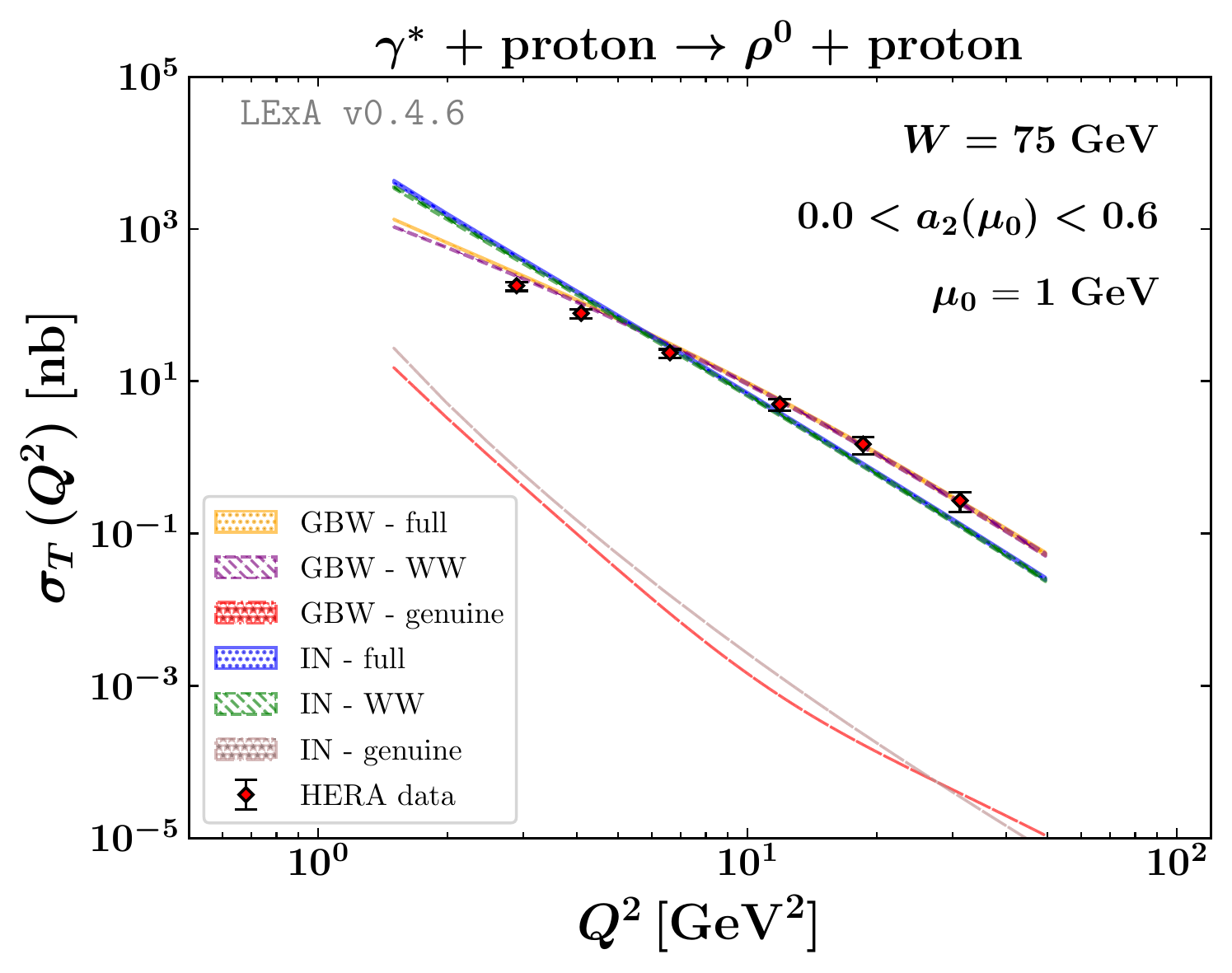}
\includegraphics[scale=0.55,clip]{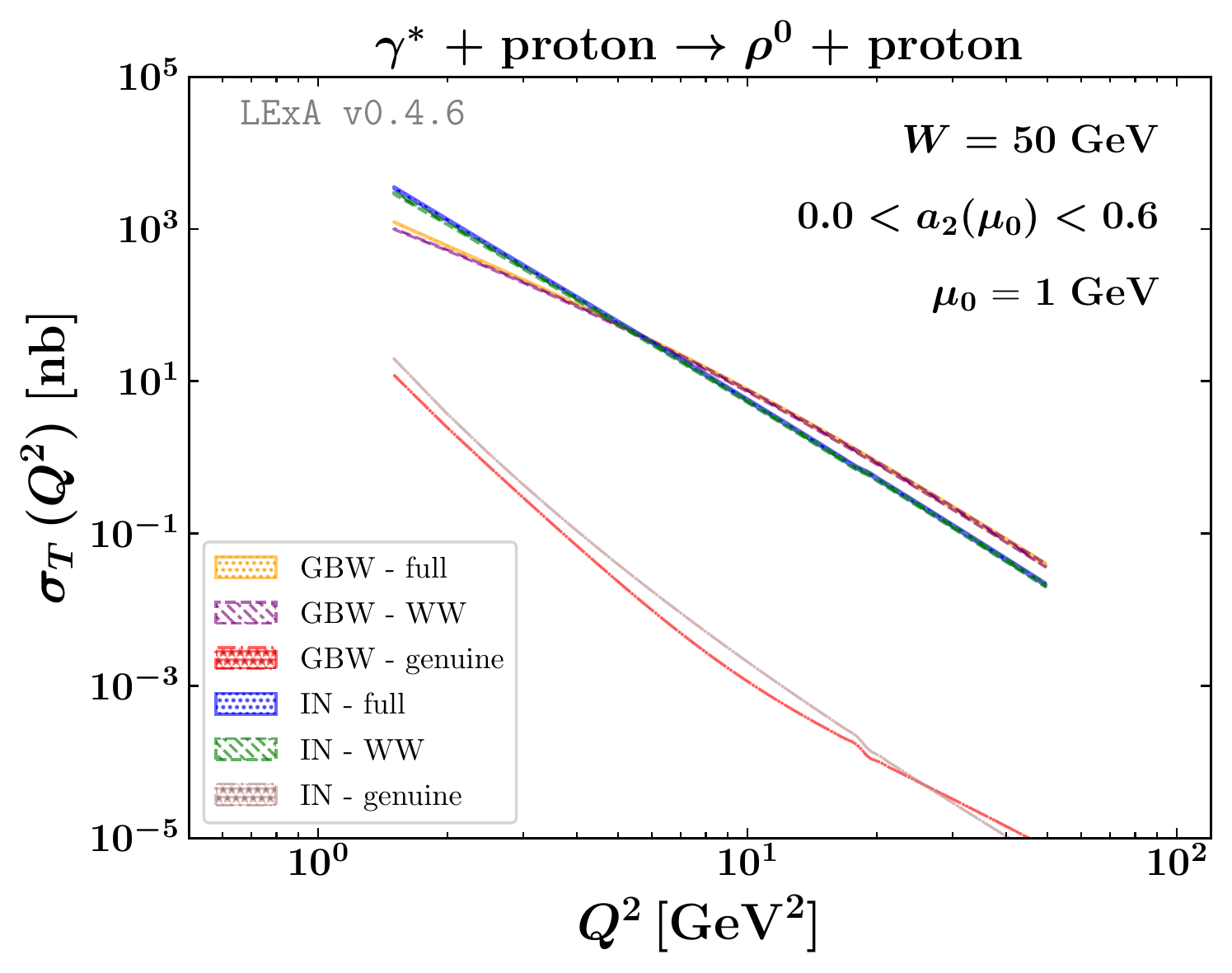}
\\
\includegraphics[scale=0.55,clip]{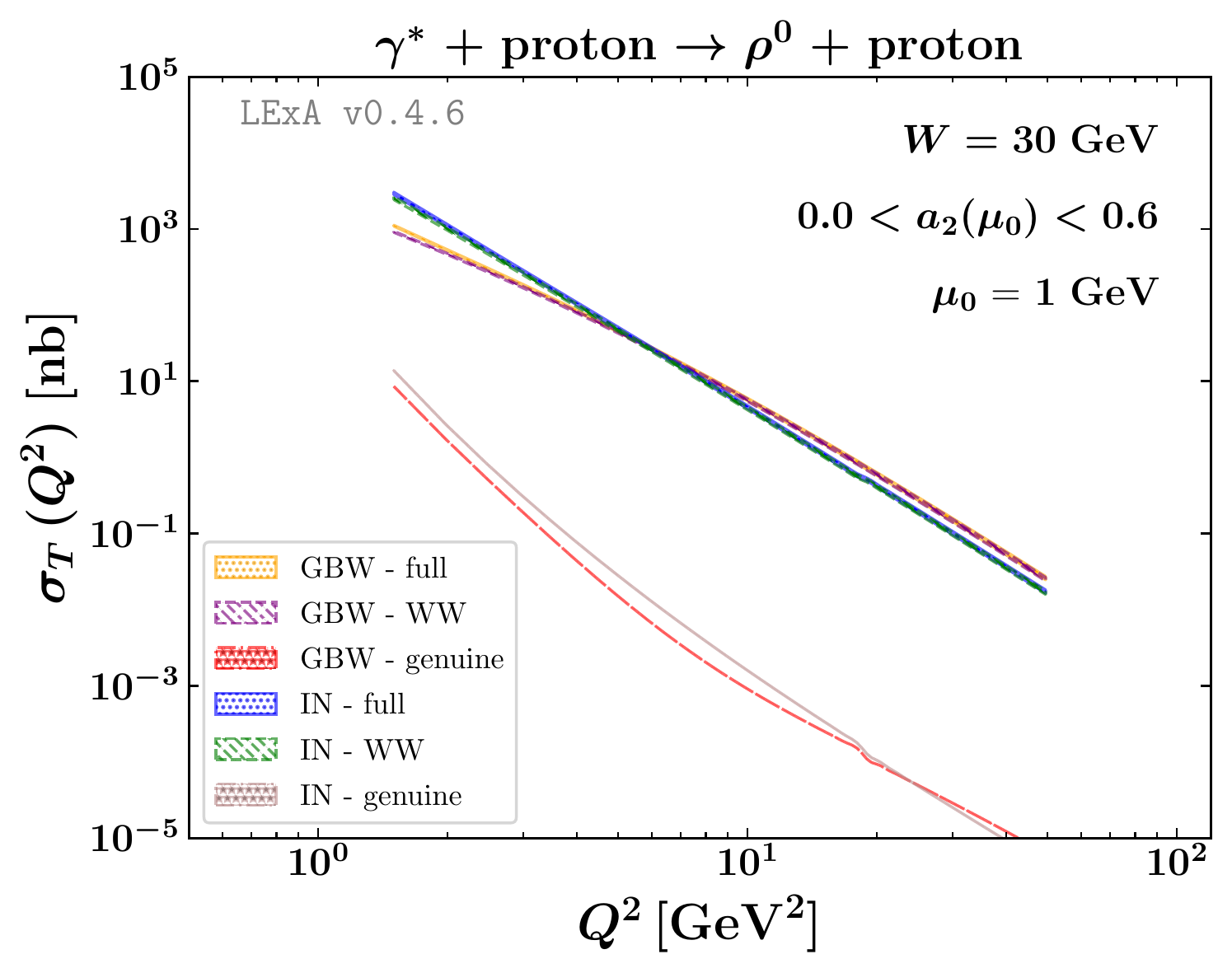}
\includegraphics[scale=0.55,clip]{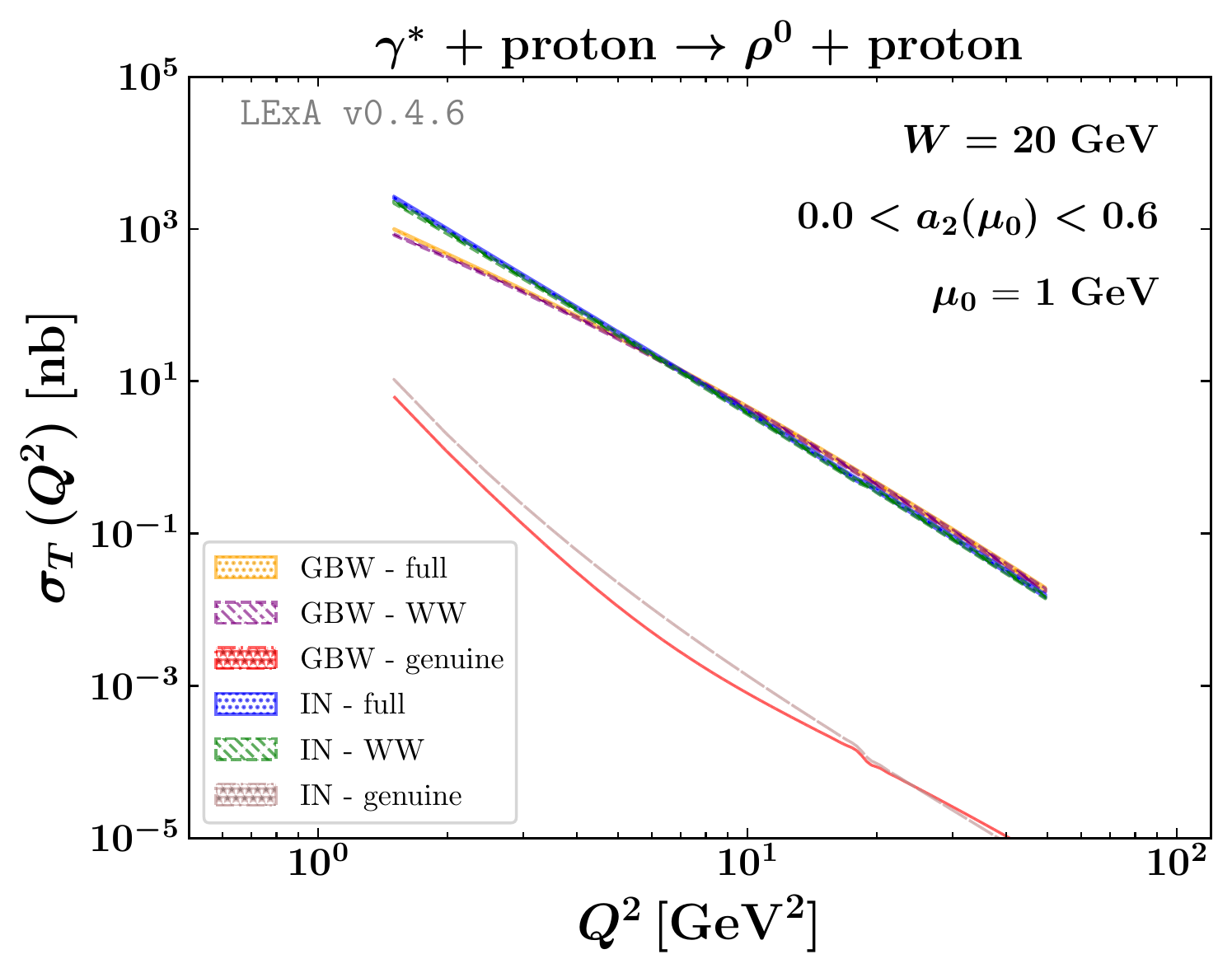}
\caption{$Q^2$-dependence of the transversely polarized cross section,
  $\sigma_T$, for GBW and IN UGD models, at $W = 75$ GeV together with
  the HERA data (left upper panel) and at $W = 20, 30, 50$ GeV relevant for EIC
  (the remaining panels). Uncertainty bands give the effect of varying
  $a_2(\mu_0 = 1\,$\rm GeV$)$ between $0.0$ and $0.6$. Full, WW and 
  genuine contributions are shown separately.}
\label{fig:sigma_T_GBW}
\end{figure}

\begin{figure}
\centering

\includegraphics[scale=0.55,clip]{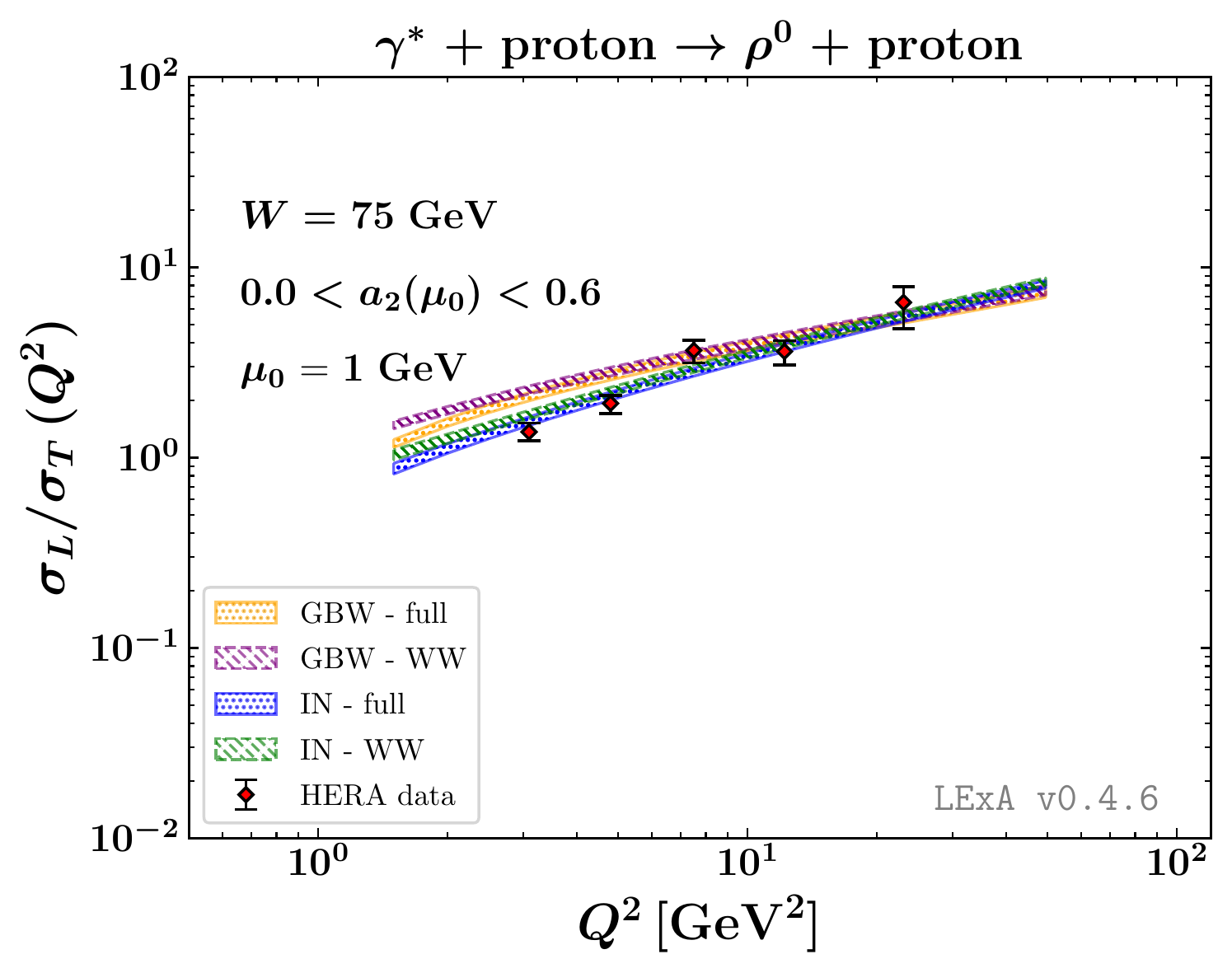}
\includegraphics[scale=0.55,clip]{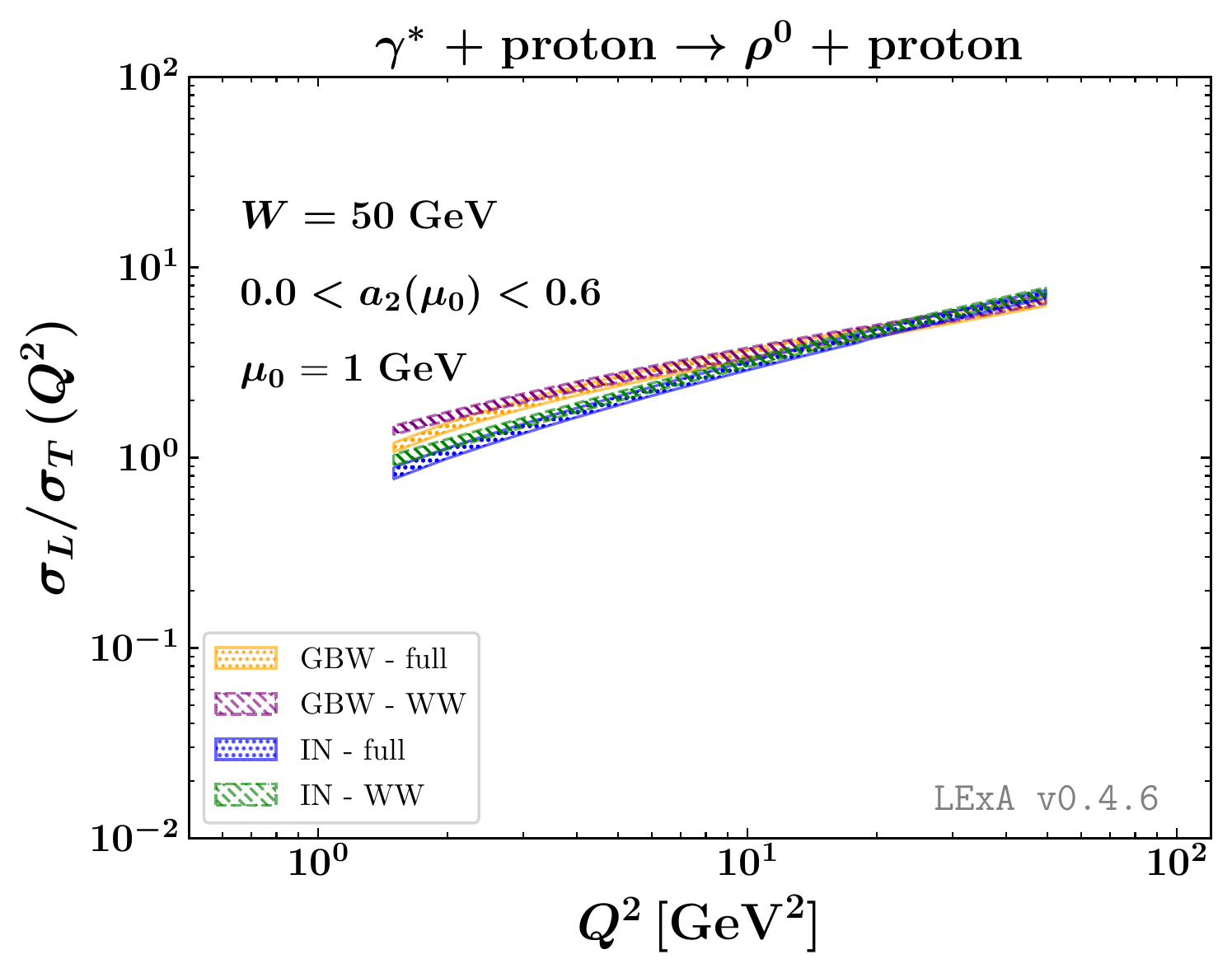}
\\
\includegraphics[scale=0.55,clip]{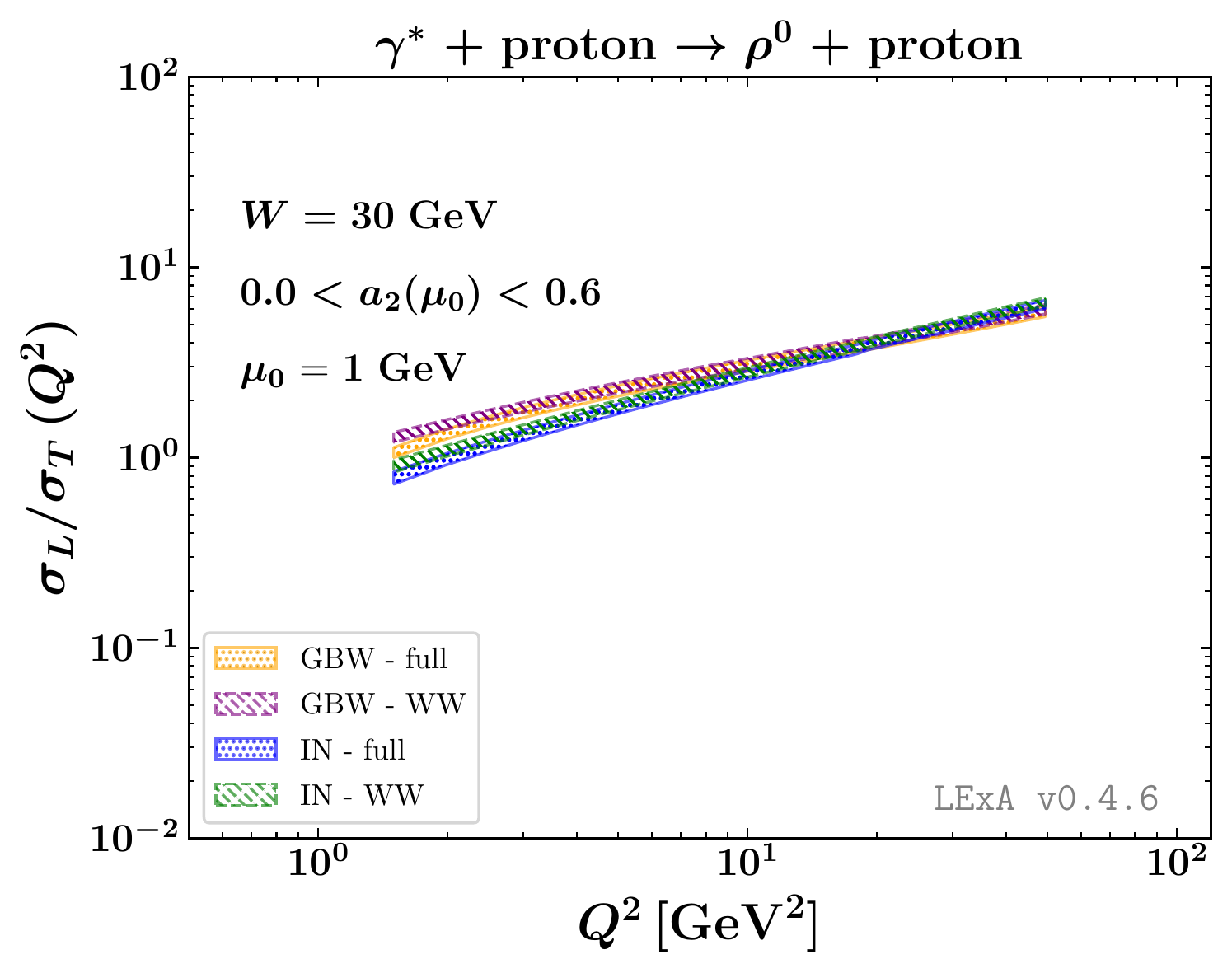}
\includegraphics[scale=0.55,clip]{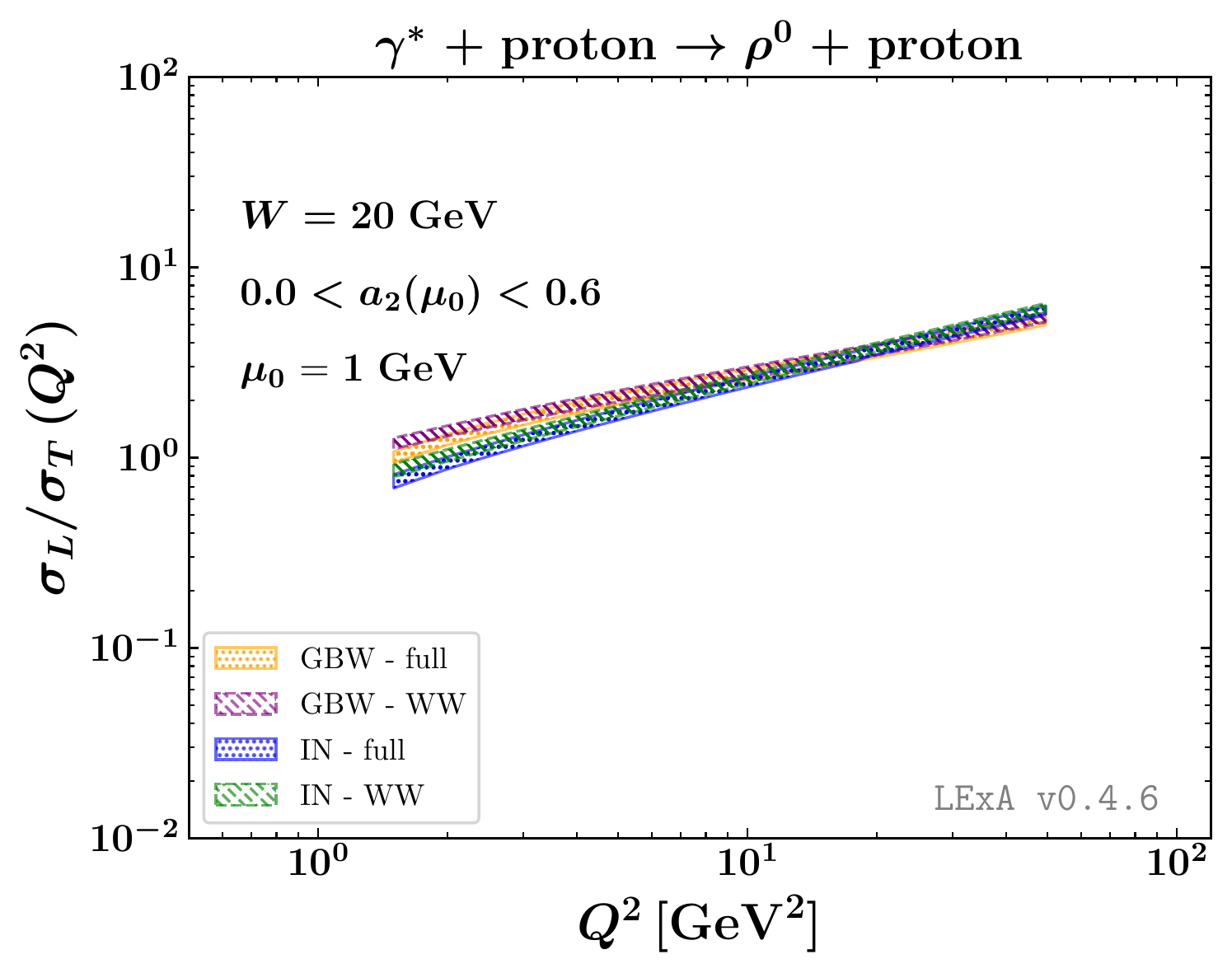}
\caption{$Q^2$-dependence of the polarized cross-section ratio,
  $\sigma_L/\sigma_T$, for the GBW and IN UGD models, at $W = 75$ GeV
  HERA (left upper panel) and at 
$W = 20, 30, 50$ GeV relevant for EIC (the remaining
panels). Uncertainty bands represent the effect of varying $a_2(\mu_0 =
1\,$\rm GeV$)$ between $0.0$ and $0.6$. Full and WW contributions are
shown separately.}
\label{fig:sigma_R_GBW}
\end{figure}

We present predictions for the polarized cross sections $\sigma_L$ and $\sigma_T$ and their ratio $\sigma_L/\sigma_T$, as obtained with all the UGD models presented above, and compare them with HERA data.
The behavior of the polarized cross sections $\sigma_L$ and $\sigma_T$
and the ratio $\sigma_L/\sigma_T$~\cite{Bolognino_PhDThesis} in terms of
$Q^2$ for all UGDs, at $W = 75$ GeV (comparison with HERA data) and at 
$W = 20, 30, 50$ GeV (predictions for the EIC), is shown in
Figs.~\ref{fig:sigma_L_all}, 
\ref{fig:sigma_T_all}, and~\ref{fig:sigma_R_all}, taking into account
the variation of the Gegenbauer coefficient $a_2(\mu_0)$ in the same
range used for the helicity-amplitude ratio analysis of Ref.\tcite{Bolognino:2018rhb}.
As can be seen from the figures, the uncertainties due to the choice of
UGDs are much bigger than those due to $a_2$ uncertainty which opens a
possibility to test UGD models for the reaction under consideration.
The comparison exhibits a
partial agreement with the experimental data, where once again none of
the proposed models is able to describe the whole $Q^2$ region. However, we can specify which UGD model is more suitable for the description of
the polarized cross sections and which, indeed, for their ratio in the
$Q^2$ intermediate range. On one side, observing
Figs.~\ref{fig:sigma_L_GBW} and~\ref{fig:sigma_T_GBW}, the GBW model, considered in its \emph{standard} definition, \emph{i.e.} without the evolution of saturation scale (for a detailed discussion about saturation effects see Ref.~\cite{Besse:2013muy}), and the
IN one appear to be the UGD models that allow us to match data for the single
polarized cross sections, $\sigma_L$ and $\sigma_T$, in the most
accurate way. Here the genuine twist-3 contribution is considered. On
the other side, although the predictions for the cross section ratio
$\sigma_L/\sigma_T$ in Fig.~\ref{fig:sigma_R_GBW} with the GBW model is
quite resonable, if we regard increasing values of $Q^2 \sim 10$ GeV$^2$,
the IN UGD model\footnote{Regarding the helicity amplitude ratio
  $T_{11}/T_{00}$, as a result of 
a correction in the numerical implementation, we note a significantly
improved agreement between 
the IN model and the experimental data with respect to the Fig.~2 in
Ref.~\cite{Bolognino:2018rhb}. 
For the sake of simplicity and recognizing in the GBW model intriguing
features and parameters worthy to be probed further, in that paper an exhaustive phenomenological analysis of this UGD parametrization was proposed.} is able to slightly better catch also the low-$Q^2$ region of data.
We stress, however, that in this low-$Q^2$ range the validity of our approach to IFs, based on the use of collinear DAs, could be questionable. Therefore our ability to discriminate among UGD models at small virtualities could be limited.

%--------------------------
\section{Conclusions}
%--------------------------
\label{conclusions}

We have calculated cross sections for the diffractive electroproduction of  $\rho$ mesons in the energy range of HERA and EIC. We have used the high-energy factorization formalism for the forward amplitude~\cite{Anikin:2009bf,Anikin:2011sa}, utilizing an empirical parametrization of the diffractive slope to obtain the relevant integrated cross sections.

The impact factors for longitudinal and transverse mesons probe the transverse 
momentum dependence of the UGD in different ways, so that
the polarization dependence of $\rho$-production has been proposed as
a sensitive probe of the shape of the UGD~\cite{Bolognino:2018mlw,Bolognino:2018rhb}. The impact factor for longitudinally polarized meson is obtained in terms of the
leading twist DA. We find that, for the
higher-twist DAs relevant for transverse mesons, the WW contributions dominate.
We have investigated the effect of uncertainty related with the form of leading twist DA by varying the Gegenbauer coefficient $a_2$.

We have performed calculations for a representative choices of UGDs
available in the literature. 
These UGDs follow different strategies in their construction.
Some explicitly connect to solutions of a specific evolution equation~\cite{Watt:2003mx,Hentschinski:2012kr}, while others stress the presence of 
a substantial nonperturbative component~\cite{GolecBiernat:1998js,Ivanov:2000cm}, 
while being adjusted to giving a 
good description of proton deep inelastic structure functions.

The latter two UGDs in fact do give the best description of the HERA cross sections. 
While our use of an empirical parametrization of the diffractive slope
introduces an additional model element/uncertainty, the spread of
predictions from various UGDs appears to be larger than the uncertainty
in the slope. Some UGDs substantially underestimate the total cross
section, which may mean that, in the relevant $\kappa$-range, their $\kappa$-shape and normalization miss some important insights on the nonperturbative proton dynamics.

The cross section ratio $\sigma_L/\sigma_T$ indeed appears to have
potential to discriminate further between UGDs, for the case at hand the 
IN UGD gives the best description of HERA data.

New data taken at the Electron-Ion Collider EIC also have a potential to
further discriminate between the different UGDs. 
At the lower range of $W$ an investigation of the
matching/correspondence 
to TMD factorization approaches with on-shell partons would be interesting.

It has been recently pointed out how the inclusive diffractive tagging of particles with a heavy transverse mass, such as Higgs bosons~\cite{Celiberto:2020tmb}, heavy-flavored jets~\cite{DafneBolognino:2019ccd,Bolognino:2019yls,Bolognino:2021mrc} and charmed baryons~\cite{Celiberto:2021dzy}, leads to a fair stabilization of the high-energy series under higher-order corrections.
Future studies of reactions featuring the forward emission of those objects will provide us with additional and possibly clearer probe channels for the UGD. Furthermore, the detection of forward quarkonia (recently studied in the context of high-energy factorization in more central directions of rapidity~\cite{Kniehl:2016sap,Cisek:2017ikn,Maciula:2018bex,Babiarz:2019mag,Babiarz:2020jkh,Babiarz:2020jhy}) in the low$-p_T$ region will help us \emph{i}) to shed light on the quarkonium production mechanisms and \emph{ii}) to investigate kinematic regions at the frontier between the high-energy and the TMD dynamics. 

The analysis presented in this work is at leading order and an obvious improvement would be its extension to the next-to-leading order. This calls for the setup of a theoretical scheme for the convolution of NLO IFs and UGD, which is not trivial, except for UGD models based on BFKL.

%----------------------------------
\section*{Acknowledgements}
%----------------------------------

A.D.B. and A.P. acknowledge support from the INFN/QFT@COLLIDERS project.
F.G.C. acknowledges support from the Italian Ministry of Education, Universities and Research under the FARE grant ``3DGLUE'' (n. R16XKPHL3N), and from the INFN/NINPHA project.
F.G.C. thanks the Universit\`a degli Studi di Pavia for the warm hospitality.
The work of D.I. was carried out within the framework of the state contract of the Sobolev Institute of Mathematics (Project No. 0314-2019-0021).
This study was partially supported by the Polish National Science Center Grant No. UMO-2018/31/B/ST2/03537 and by the Center for Innovation and Transfer of Natural Sciences and Engineering Knowledge in Rzesz\'ow.

The diagram in Fig.~\ref{fig:process_rho} was realized via the {\tt JaxoDraw 2.0} interface~\cite{Binosi:2008ig}.
%\newpage

\bibliographystyle{apsrev}
\bibliography{references}

\end{document}